\documentclass[aip,jap,amsmath,10pt,twocolumn,superscriptaddress]{revtex4-1}
\usepackage[dvipdfmx]{graphicx}
\usepackage{color}
\usepackage{bm}
\usepackage{braket}
\usepackage{amsmath,amssymb}

\begin{document}
\title{Systematic development of polynomial machine learning potentials for metallic and alloy systems}
\author{Atsuto \surname{Seko}}
\email{seko@cms.mtl.kyoto-u.ac.jp}
\affiliation{Department of Materials Science and Engineering, Kyoto University, Kyoto 606-8501, Japan}
\date{\today}

\begin{abstract}
Machine learning potentials (MLPs) developed from extensive datasets constructed from density functional theory (DFT) calculations have become increasingly appealing for many researchers. 
This paper presents a framework of polynomial-based MLPs, called polynomial MLPs.
The systematic development of accurate and computationally efficient polynomial MLPs for many elemental and binary alloy systems and their predictive powers for various properties are also demonstrated.
Consequently, many polynomial MLPs are available in a repository website \cite{MachineLearningPotentialRepository}.
The repository will help many scientists perform accurate and efficient large-scale atomistic simulations and crystal structure searches.
\end{abstract}

\maketitle

\section{Introduction}

Machine learning potentials (MLPs) have been increasingly required to perform crystal structure optimizations and large-scale atomistic simulations more accurately than with conventional interatomic potentials.
MLPs have been generally developed from large datasets constructed by systematic density functional theory (DFT) calculations; hence, they can improve the accuracy and transferability of interatomic potentials significantly.
Therefore, many recent studies have proposed a number of procedures to develop MLPs and have shown their applications
\cite{
Lorenz2004210,
behler2007generalized,
bartok2010gaussian,
behler2011atom,
han2017deep,
258c531ae5de4f5699e2eec2de51c84f,
PhysRevB.96.014112,
PhysRevB.90.104108,
PhysRevX.8.041048,
PhysRevLett.114.096405,
PhysRevB.95.214302,
PhysRevB.90.024101,
PhysRevB.92.054113,
PhysRevMaterials.1.063801,
Thompson2015316,
wood2018extending,
PhysRevMaterials.1.043603,
doi-10.1137-15M1054183,
PhysRevLett.120.156001,
podryabinkin2018accelerating,
GUBAEV2019148,
doi:10.1063/1.5126336,
Freitas2022}.
Simultaneously, MLPs themselves are necessary for their users to perform accurate atomistic simulations. 
Therefore, the development and release of MLPs for a wide range of systems should be useful, similarly to the conventional interatomic potentials available in several repositories \cite{interatomicPotentialRepository,KIMproject}.

This paper reviews a formulation representing the potential energy in elemental and alloy systems. 
In the formulation, polynomial models of polynomial invariants for the O(3) group, which are referred to as ``polynomial MLPs'', are derived. 
This paper also includes procedures for generating structure datasets used for the DFT calculation, estimating the polynomial MLPs, and finding the optimal MLPs.
Because the accuracy and computational efficiency of polynomial MLPs are conflicting properties, there is no single optimal MLP, and the solutions are the Pareto-optimal MLPs with different trade-offs between them\cite{PhysRevB.99.214108,hernandez2019fast,doi:10.1021/acs.jpca.9b08723}.
The accuracy and predictive power of polynomial MLPs are also demonstrated for elemental and binary alloy systems.

The Pareto-optimal polynomial MLPs developed using the current procedures are available in the \textsc{Polynomial Machine Learning Potential Repository} \cite{MachineLearningPotentialRepository}, in which many physical properties predicted using the polynomial MLPs can be found.
Moreover, a user package that combines the polynomial MLPs with atomistic simulations using the \textsc{lammps} code \cite{lammps} is also available on a website \cite{LammpsPolyMLP}.
The polynomial MLPs for 50 elemental and 114 binary alloy systems, developed from DFT calculations for approximately two million structures, are now available.

Section \ref{tutorial-2022:Sec-method} introduces the formulation of the potential energy models and procedures for developing the polynomial MLPs.
In Sec. \ref{tutorial-2022:Sec-mlp-repository}, the Pareto-optimal polynomial MLPs for a wide variety of elemental and binary alloy systems are shown. 
A procedure for selecting an appropriate MLP from the whole set of the Pareto-optimal MLPs is also proposed.
The selection of MLPs is an essential process for performing the subsequent large-scale atomistic simulations.
In Sec. \ref{tutorial-2022:Sec-predictive-power}, the predictive powers of the Pareto-optimal MLPs for various properties are demonstrated.
Finally, this study is summarized in Sec. \ref{tutorial-2022:Sec-conclusion}.

\section{Methodology}
\label{tutorial-2022:Sec-method}

This section introduces potential energy models used for developing the polynomial MLPs and procedures for generating DFT datasets and estimating potential energy models.
Section \ref{tutorial-2022:Sec-method-formulation} derives a general formulation of the relationship between the potential energy and the neighboring atomic distribution as a starting point of developing polynomial MLPs and other types of MLPs.
Section \ref{tutorial-2022:Sec-method-structural-features} shows polynomial invariants composed of order parameters representing the neighboring atomic distribution.
The current polynomial invariants are systematically enumerated from order parameters in terms of products of radial and spherical harmonic functions.
Section \ref{tutorial-2022:Sec-models} suggests polynomial functions representing the potential energy with respect to the polynomial invariants.
Section \ref{tutorial-2022:Sec-method-dataset} explains about the DFT datasets required to estimate the polynomial MLPs. 
The current datasets are generated so that the polynomial MLPs have high predictive power for a wide variety of structures.
Section \ref{tutorial-2022:Sec-regression} shows computational procedures for estimating the coefficients of the polynomial MLPs.
In Sec. \ref{tutorial-2022:Sec-method-pareto-optimality}, a procedure for finding the optimal MLPs is demonstrated.
Although the potential energy will be formulated for multicomponent systems throughout this study, element indices can be ignored in the formulation for elemental systems.

\subsection{A general formulation of potential energy}
\label{tutorial-2022:Sec-method-formulation}

\begin{figure}[tbp]
\includegraphics[clip,width=\linewidth]{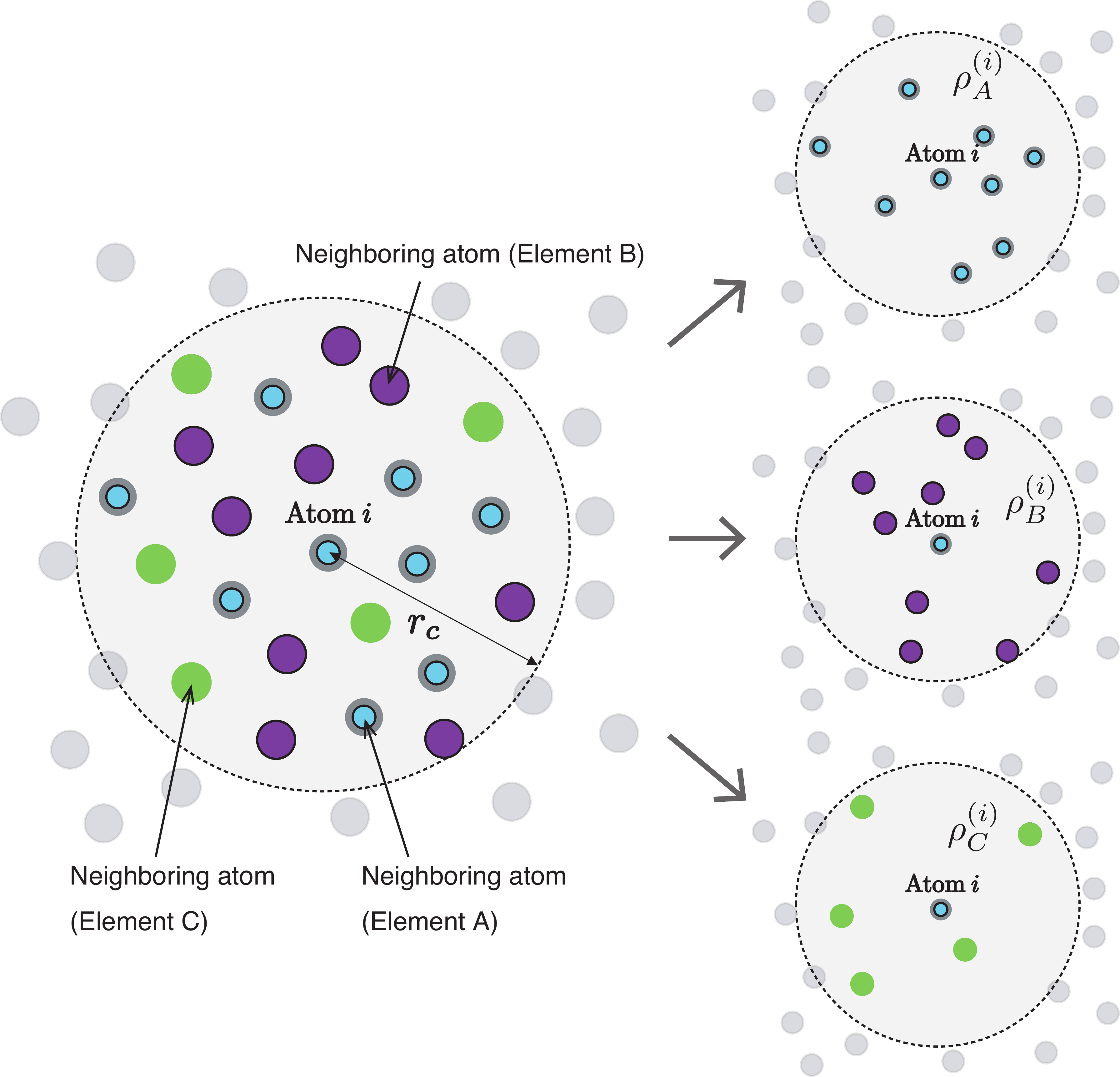}
\caption{
Schematic illustration of the neighboring atomic distribution around atom $i$ of element A in an A-B-C ternary structure and its decomposition into the neighboring atomic densities of elements A, B, and C around atom $i$ of element A.
}
\label{tutorial-2022:atomic-distribution-schematic}
\end{figure}

The short-range part of the potential energy for a structure, $E$, is assumed to be decomposed as
\begin{equation}
E = \sum_i E^{(i)},
\end{equation}
where $E^{(i)}$ denotes the contribution of atom $i$ within a given cutoff radius $r_c$, referred to as the atomic energy.
The atomic energy of atom $i$ is then approximately given by a functional of its neighboring atomic densities.
Figure \ref{tutorial-2022:atomic-distribution-schematic} shows a schematic illustration of the neighboring atomic distribution around atom $i$ within cutoff radius $r_c$ and its decomposition into the neighboring atomic densities of elements.
In a multicomponent system composed of elements $\{\rm A, \rm B, \cdots\}$, the atomic energy is written using functional $\mathcal{F}$ dependent on the element of atom $i$ as
\begin{equation}
E^{(i)} = \mathcal{F}_{s_i} \left[ \rho^{(i)}_{(s_i, \rm A)} , \rho^{(i)}_{(s_i, \rm B)}, \cdots \right],
\end{equation}
where $\rho^{(i)}_{(s_i,s)}$ denotes the neighboring atomic density of element $s$ ($s \in \{{\rm A},{\rm B},\cdots\}$) around atom $i$ of element $s_i$.

The neighboring atomic density of element $s$ around atom $i$ is then expanded in terms of a basis set $\{b_n\}$. 
The expansion enables the functional form to be replaced with a function of its expansion coefficients.
The neighboring atomic densities can be expanded as
\begin{equation}
\rho^{(i)}_{(s_i, s)}(\bm{r}) = \sum_n a_{n,(s_i, s)}^{(i)} b_n(\bm{r}),
\end{equation}
where $a_{n,(s_i,s)}^{(i)}$ denotes the $n$th order parameter characterizing the neighboring atomic density of element $s$ around atom $i$ of element $s_i$.
By introducing order parameters for unordered pairs of elements, $a_{n,t}^{(i)}$, where $t \in \{ \{\rm A,\rm A\}, \{\rm A,\rm B\}, \cdots\}$, and defining order parameter $a_{n,t}^{(i)}$ to be zero if $s_i$ is not included in $t$, the atomic energy is written as
\begin{equation}
\label{tutorial-2022:Eqn-atomic-energy-ops}
E^{(i)} = F' \left(a_{1,t_1}^{(i)}, a_{1,t_2}^{(i)}, 
\cdots, a_{2,t_1}^{(i)}, a_{2,t_2}^{(i)}, \cdots\right).
\end{equation}
In this equation, all pair combinations with the replacement of elements are considered for each $n$.
A more detailed derivation of Eqn. (\ref{tutorial-2022:Eqn-atomic-energy-ops}) is found in Ref. \onlinecite{PhysRevB.102.174104}.

Moreover, an arbitrary rotation leaves the atomic energy invariant, although it generally changes the neighboring atomic densities and their order parameters\cite{PhysRevB.99.214108}.
Therefore, the atomic energy should be a function of O(3) invariants $\{d_{m'}^{(i)}\}$ derived from the order parameters $\{a_{n,t}^{(i)}\}$, expressed as
\begin{equation}
\label{tutorial-2022:Eqn-atomic-energy-features}
E^{(i)} = F \left( d_1^{(i)}, d_2^{(i)}, \cdots \right).
\end{equation}
A number of functions are useful as function $F$ to represent the relationship between the invariants and the atomic energy, such as
artificial neural network models
\cite{
Lorenz2004210,
behler2007generalized,
behler2011atom,
han2017deep,
258c531ae5de4f5699e2eec2de51c84f,
PhysRevB.96.014112}, 
Gaussian process models
\cite{
bartok2010gaussian,
PhysRevB.90.104108,
PhysRevX.8.041048,
PhysRevLett.114.096405,
PhysRevB.95.214302},
and linear models
\cite{
PhysRevB.90.024101,
PhysRevB.92.054113,
PhysRevMaterials.1.063801,
Thompson2015316,
wood2018extending,
PhysRevMaterials.1.043603,
doi-10.1137-15M1054183}.
The current formulation is useful as a starting point for deriving new potential energy models.

\subsection{Structural features}
\label{tutorial-2022:Sec-method-structural-features}

The invariants composed of order parameters representing the neighboring atomic density are hereafter referred to as ``structural features'' for representing potential energy.
A procedure to systematically enumerate structural features that can control the accuracy and computational efficiency of MLPs (e.g., Refs. \onlinecite{bartok2013representing,PhysRevB.99.214108}) plays an essential role in automatically developing accurate and efficient MLPs. 
Therefore, the current MLP employs systematic sets of polynomial invariants up to the sixth order derived from order parameters representing the neighboring atomic density in terms of radial and spherical harmonic functions.

When the neighboring atomic density is expanded in terms of products of radial functions $\{f_n\}$ and spherical harmonics $\{Y_{lm}\}$, the neighboring atomic density of element $s$ at a position $(r, \theta, \phi)$ in spherical coordinates centered at the position of atom $i$ is expressed as
\begin{equation}
\rho^{(i)}_{(s_i,s)} (r, \theta, \phi) = \sum_{nlm} a^{(i)}_{nlm,\{s_i,s\}} f_n(r) Y_{lm} (\theta, \phi),
\end{equation}
where order parameter $a^{(i)}_{nlm,\{s_i,s\}}$ is component $nlm$ of the neighboring atomic density of atom $i$.
Although the order parameters are not generally invariant for the O(3) group, 
a $p$th-order polynomial invariant for a radial index $n$ and a set of pairs composed of the angular number and the element unordered pair $\{(l_1,t_1),(l_2,t_2),\cdots,(l_p,t_p)\}$ is given by a linear combination of products of $p$ order parameters, expressed as
\begin{widetext}
\begin{equation}
\label{tutorial-2022:Eqn-invariant-form}
d_{nl_1l_2\cdots l_p,t_1t_2\cdots t_p,(\sigma)}^{(i)} =
\sum_{m_1,m_2,\cdots, m_p} c^{l_1l_2\cdots l_p,(\sigma)}_{m_1m_2\cdots m_p}
a_{nl_1m_1,t_1}^{(i)} a_{nl_2m_2,t_2}^{(i)} \cdots a_{nl_pm_p,t_p}^{(i)},
\end{equation}
\end{widetext}
where coefficient set $\{c^{l_1l_2\cdots l_p,(\sigma)}_{m_1m_2\cdots m_p}\}$ is independent of the radial index $n$ and the element unordered pair $t$. 

The coefficient set ensures that the linear combinations are invariant for arbitrary rotation.
A numerical way to obtain linearly independent coefficient sets for a given set $\{l_1,l_2,\cdots,l_p\}$ is the group-theoretical projector operation method, which is the general process of reducing the Kronecker products of irreducible representations \cite{el-batanouny_wooten_2008,PhysRevB.99.214108}.
The reduction of Kronecker products of irreducible representations has been widely used for many purposes in physics and chemistry, such as the formulation of angular momentum coupling, the derivation of selection rules, and the formulation of the Landau free energy for phase transitions\cite{el-batanouny_wooten_2008,1987ltpt.book,Hamermesh:1123140,heine2007group,chaichian1997symmetries}.
In terms of fourth- and higher-order polynomial invariants, multiple invariants are linearly independent for most of the set $\{l_1,l_2,\cdots,l_p\}$, which are distinguished by index $\sigma$ if necessary.

\begin{figure}[tbp]
\includegraphics[clip,width=0.8\linewidth]{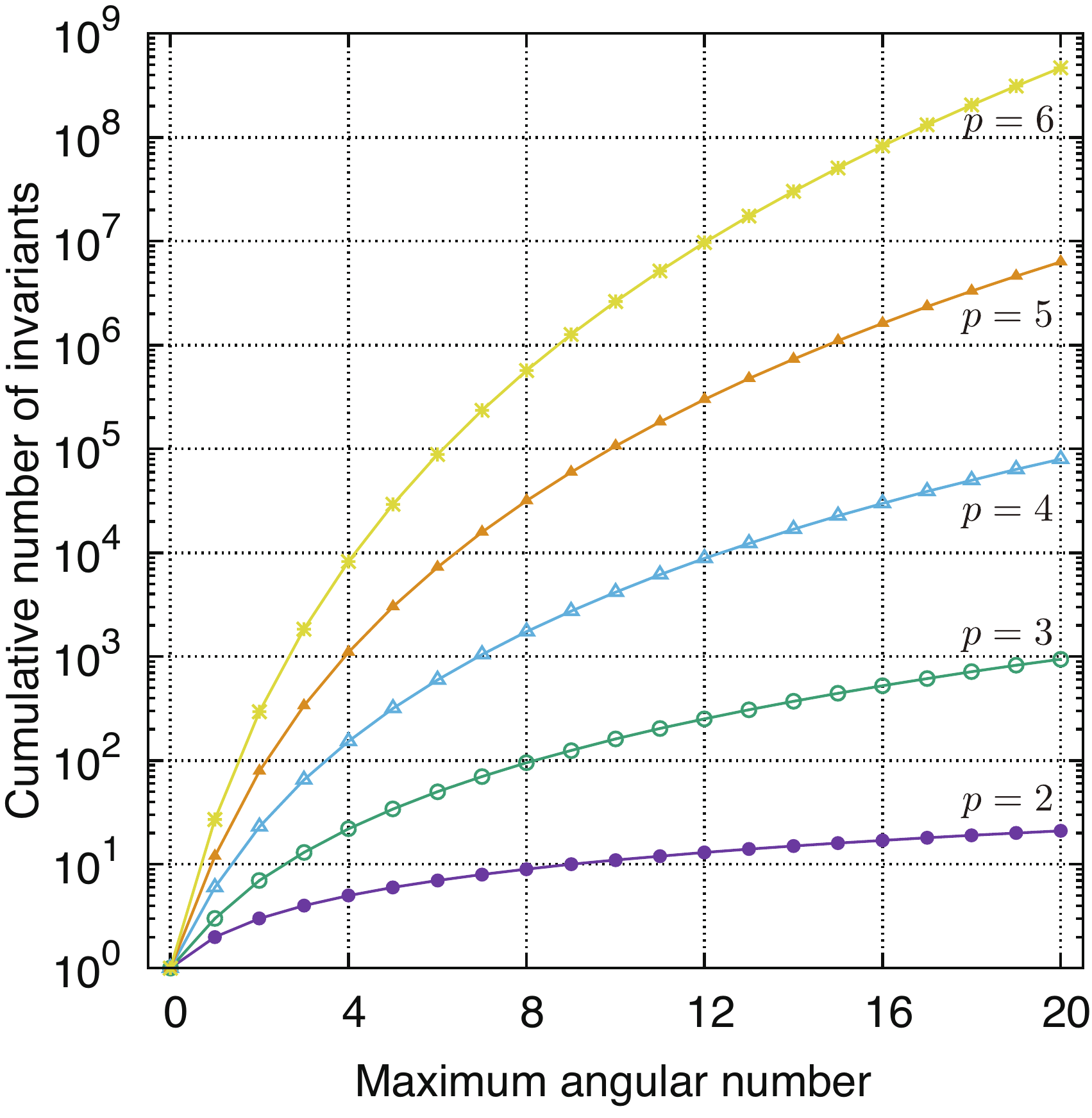}
\caption{
Cumulative number of SO(3) invariants for a given value of maximum $l$.
}
\label{tutorial-2022:Fig-number-invariants}
\end{figure}

Figure \ref{tutorial-2022:Fig-number-invariants} shows the number of $p$th-order invariants of the SO(3) group satisfying $l_1 \leq l_{\rm{max}}$, $l_2 \leq l_{\rm{max}}$, $\cdots$, $l_p \leq l_{\rm{max}}$ for a given $l_{\rm{max}}$.
The integer sequences for second- and third-order invariants may correspond to the On-Line Encyclopedia of Integer Sequences (OEISs) A000027 and A002623, respectively\cite{oeis}.
Among these invariants, only polynomials that are invariant for an arbitrary improper rotation are adopted.
They correspond to invariants of the O(3) group and are polynomial invariants for the sets of $\{l_1,l_2,\cdots,l_p\}$ whose sums $l_1+l_2+\cdots+l_p$ are even.
In addition, these numbers include linearly dependent and constantly zero invariants when considering products derived from a single basis set.
Therefore, the number of available invariants is obtained by solving the eigenvalue problems for the projector matrix and removing such invariants \cite{PhysRevB.99.214108}.

Note that the second- and third-order invariants are equivalent to a multicomponent extension of the angular Fourier series and the bispectrum reported in the literature, respectively \cite{kondor2008group,bartok2013representing}.
When excluding radial parts, the second-order invariants are exactly the same as the bond-orientational order parameters (BOPs)\cite{PhysRevB.28.784}.
If we restrict the invariants to third-order symmetrized ones excluding radial parts, they are third-order BOPs.

Moreover, nonzero polynomial invariants are derived from the sets of $t_p$ satisfying the condition that the intersection of $t_p$ is not the empty set,
\begin{equation}
\prod_p t_p \neq \varnothing.
\end{equation}
An example where the intersection of element pairs becomes the empty set is the case that a polynomial invariant is composed of order parameters with $t_1 = {\rm \{A,A\}}$ and $t_2 = {\rm \{B,B\}}$.
Such polynomial invariants are eliminated.
Using this condition, possible sets of pairs that are composed of the angular number and the element unordered pair $\{(l_1,t_1),(l_2,t_2),\cdots,(l_p,t_p)\}$ are enumerated for a given maximum angular number to obtain the entire set of polynomial invariants.

The current MLPs adopt a finite set of Gaussian-type radial functions as $f_n$ given by
\begin{equation}
f_{n}(r)=\exp\left[-\beta_n(r-r_n)^{2}\right] f_c(r),
\end{equation}
where $\beta_n$ and $r_n$ denote parameters.
Cutoff function $f_c$ ensures the smooth decay of the radial function, and the current MLP employs a cosine-based cutoff function expressed as 
\begin{eqnarray}
f_c(r) = \left\{
\begin{aligned}
& \frac{1}{2} \left[ \cos \left( \pi \frac{r}{r_c} \right) + 1\right] & (r \le r_c)\\
& 0 & (r > r_c)
\end{aligned}
\right ..
\end{eqnarray}
The order parameters of atom $i$ and element pair $\{s_i,s\}$ are approximately estimated from the neighboring atomic density of element $s$ around atom $i$ as
\begin{equation}
a_{nlm,\{s_i,s\}}^{(i)} = \sum_{\{j | r_{ij} \leq r_c,s_j = s\} }
f_n(r_{ij}) Y_{lm}^* (\theta_{ij}, \phi_{ij}),
\end{equation}
where $(r_{ij}, \theta_{ij}, \phi_{ij})$ denotes the spherical coordinates of neighboring atom $j$ centered at the position of atom $i$.
Although the Gaussian-type radial functions are not orthonormal, such an approximation of the order parameters is acceptable in formulating the polynomial MLP, as discussed in Ref. \onlinecite{PhysRevB.99.214108}.

In the special case that the neighboring atomic density is expanded in terms of only radial functions $\{f_n\}$, the neighboring atomic density around atom $i$ is expressed as
\begin{equation}
\rho^{(i)}_{(s_i,s)} (r) = \sum_{n} a^{(i)}_{n,\{s_i,s\}} f_n(r).
\end{equation}
Since order parameter $a^{(i)}_{n,\{s_i,s\}}$ is invariant for the O(3) group, it can be a pairwise structural feature denoted as
\begin{equation}
d^{(i)}_{n0,t} = a^{(i)}_{n,t}.
\end{equation}
A pairwise structural feature is then approximately estimated as
\begin{equation}
d_{n0}^{(i)} = \sum_{j \in {\rm neighbor}} f_n(r_{ij}).
\end{equation}

\subsection{Polynomial models}
\label{tutorial-2022:Sec-models}

The current MLPs employ the following polynomial functions representing the relationship between the atomic energy and structural features as function $F$ in Eqn. (\ref{tutorial-2022:Eqn-atomic-energy-features}).
Given a set of structural features $D = \set{d_1,d_2,\cdots}$, the polynomial functions are represented as
\begin{eqnarray}
F_1 \left(D\right) &=& \sum_{i'} w_{i'} d_{i'} \nonumber \\
F_2 \left(D\right) &=& \sum_{\{i',j'\}} w_{i'j'} d_{i'} d_{j'} \\
F_3 \left(D\right) &=& \sum_{\{i',j',k'\}} w_{i'j'k'} d_{i'} d_{j'} d_{k'} \nonumber \\
& \vdots & \nonumber
\end{eqnarray}
where $w$ denotes a regression coefficient.
Polynomial function $F_\xi$ is composed of all combinations of $\xi$ structural features. 

A polynomial MLP in the repository is identified with each of the following combinations of the polynomial functions and structural features.
The current models have no constant term; hence, the atomic energy is measured from the sum of the energies of isolated atoms.
The simplest model is a polynomial of pairwise structural features.
When a set of pairwise structural features is described by
\begin{equation}
D_{\rm pair}^{(i)} = \set{d_{n0,t}^{(i)}},
\end{equation}
the pairwise polynomial model is expressed as
\begin{equation}
E^{(i)} = F_1\left(D_{\rm pair}^{(i)} \right)
+ F_2\left(D_{\rm pair}^{(i)} \right)
+ F_3\left(D_{\rm pair}^{(i)} \right)
+ \cdots.
\end{equation}
The pairwise polynomial model includes the special case that only powers of the pairwise structural features are considered, which was introduced for elemental systems in Refs. \onlinecite{PhysRevB.90.024101} and \onlinecite{PhysRevB.92.054113}.

Because the current MLPs are regarded as extensions of conventional interatomic potentials, a classification rule of conventional interatomic potentials based on the type of structural features\cite{carlsson1990beyond} is applicable to the current MLPs.
Following the classification in Ref. \onlinecite{carlsson1990beyond}, the pairwise polynomial model is classified into a pair functional potential. 
In addition, as discussed in Ref. \onlinecite{PhysRevMaterials.1.063801}, the pairwise polynomial model can also be regarded as a straightforward extension of embedded atom method (EAM) potentials.

The second model is the linear polynomial of polynomial invariants given by Eqn. (\ref{tutorial-2022:Eqn-invariant-form}).
The second model is written as
\begin{equation}
E^{(i)} = F_1 \left( D^{(i)} \right),
\label{Eqn-linear-polynomial}
\end{equation}
which was introduced in Ref. \onlinecite{PhysRevB.99.214108} for elemental systems.
A set of polynomial invariants $D^{(i)}$ is described as
\begin{equation}
D^{(i)} = D_{\rm pair}^{(i)} \cup D_2^{(i)}
\cup D_3^{(i)} \cup D_4^{(i)} \cup \cdots,
\end{equation}
where a set of $p$th-order polynomial invariants is denoted by
\begin{eqnarray}
D_2^{(i)} &=& \set{d_{nll,t_1t_2}^{(i)}} \nonumber \\
D_3^{(i)} &=& \set{d_{nl_1l_2l_3,t_1t_2t_3}^{(i)}} \\
D_4^{(i)} &=& \set{d_{nl_1l_2l_3l_4,t_1t_2t_3t_4,(\sigma)}^{(i)}} \nonumber.
\end{eqnarray}
A second-order invariant is identified with a single $l$ value because second-order linear combinations are invariant only when $l_1=l_2$ \cite{el-batanouny_wooten_2008,1987ltpt.book}.
Note that a linear polynomial model with up to third-order invariants is equivalent to a spectral neighbor analysis potential (SNAP) \cite{Thompson2015316}, expressed as
\begin{equation}
E^{(i)} = F_1 \left( D_{\rm pair}^{(i)} \cup D_2^{(i)} \cup D_3^{(i)} \right).
\end{equation}

The most general model is a polynomial of polynomial invariants described as
\begin{equation}
\label{Eqn-polynomial-model1}
E^{(i)} = F_1 \left( D^{(i)} \right) + F_2 \left( D^{(i)} \right)
+ F_3 \left( D^{(i)} \right) + \cdots.
\end{equation}
It is classified as a cluster functional potential using the classification in Ref. \onlinecite{carlsson1990beyond}.
Note that a quadratic polynomial model of polynomial invariants up to the third order is equivalent to a quadratic SNAP \cite{doi:10.1063/1.5017641}.
Moreover, a polynomial model of polynomial invariants up to the second order, which are equivalent to angular structural features, is a generalization of modified EAM (MEAM) potentials \cite{PhysRevMaterials.1.063801}.

Other extended models are also introduced, which are given by
\begin{eqnarray}
\label{Eqn-polynomial-model2}
E^{(i)} &=& F_1 \left( D^{(i)} \right) + F_2 \left( D_{\rm pair}^{(i)} \right)
+ F_3 \left( D_{\rm pair}^{(i)} \right) \nonumber \\
E^{(i)} &=& F_1 \left( D^{(i)} \right)
+ F_2 \left( D_{\rm pair}^{(i)} \cup D_2^{(i)} \right) \\
E^{(i)} &=& F_1 \left( D^{(i)} \right)
+ F_2 \left( D_{\rm pair}^{(i)} \cup D_2^{(i)} \cup D_3^{(i)} \right). \nonumber
\end{eqnarray}
They are decomposed into a linear polynomial of structural features and a polynomial of a subset of the structural features.

In multicomponent systems, only nonzero polynomial terms are retained, which is analogous to the enumeration of nonzero polynomial invariants.
A structural feature is composed of order parameters, each of which has an attribute on the element unordered pair $t$.
Therefore, when the element pair of the $p'$th order parameter in structure feature $d_{i'}$ of a polynomial term is denoted by $t_{i',p'}$, a nonzero polynomial term satisfies the condition that the intersection of $\{t_{i',p'}\}$ is not the empty set:
\begin{equation}
\left( \prod_{p'} t_{i',p'} \right) \cap 
\left( \prod_{p'} t_{j',p'} \right) \cap \cdots \neq \varnothing.
\end{equation}
For example, the intersection of element pairs becomes the empty set if a polynomial term is composed of structural features with $t_{i',p_i'} = {\rm \{A,A\}}$ and $t_{j',p_j'} = {\rm \{B,B\}}$.
Such polynomial terms are eliminated from the polynomial functions.

Finally, the advantages of the polynomial MLPs will be emphasized.
The polynomial MLPs can be regarded as linear models with respect to the structural features and their products.
Therefore, their model coefficients can be efficiently estimated by using well-optimized efficient algorithms of linear regression.
In particular, when developing many MLPs for a wide variety of systems, such an efficient estimation of regression coefficients is essential.
In addition, the forces acting on atoms and the stress tensor are also expressed by linear models with the regression coefficients for the potential energy, as derived in Ref. \onlinecite{PhysRevB.99.214108}.
Therefore, the forces and the stress tensor obtained by the DFT calculation can be used for regression in a straightforward manner as will be shown in Sec. \ref{tutorial-2022:Sec-regression}.

\subsection{Datasets}
\label{tutorial-2022:Sec-method-dataset}
Each of the polynomial MLPs is developed from a training dataset, and the prediction errors for the energy, force, and stress tensor are estimated using a test dataset.
Here, training and test datasets are generated from prototype structures, whose atomic positions and lattice constants are fully optimized by the DFT calculation.
They are referred to as structure generators.

For elemental systems, structure generators are prototype structures reported in the Inorganic Crystal Structure Database (ICSD) \cite{bergerhoff1987crystal}, which aims to cover a wide variety of structures.
Among the ICSD entries, all prototype structures composed of single elements with zero oxidation state are chosen.
Therefore, the total number of structure generators is 86. 
The list of structure generators can be found in the Appendix of Ref. \onlinecite{PhysRevB.99.214108}.
For binary alloy systems, many prototype structures reported as binary alloy entries are also adopted as structure generators.
Here, the prototype structures are restricted to those represented by unit cells with up to eight atoms.
A structure made by swapping elements in each prototype structure is also considered in the set of structure generators.
The total number of binary structure generators is 150, and they are listed in the Appendix of Ref. \onlinecite{PhysRevB.102.174104}.

For each elemental system, 13000--15000 structures are generated from the structure generators using the following procedure. 
They are randomly divided into training and test datasets at the ratio of nine to one.
For each binary alloy system, 30000--50000 structures are included in either dataset, where the structures used for the elemental systems are also included.
Each structure in the datasets is constructed by introducing random lattice expansion, random lattice distortion, and random atomic displacements into a supercell of a structure generator.
When matrix $\bm{A}$ and vector $\bm{f}$ represent the lattice vectors of the original supercell and the fractional coordinates of the atom in the original supercell, respectively, the lattice vectors of the new structure $\bm{A'}$ and the fractional coordinates of an atom in the new structure $\bm{f'}$ are given as
\begin{eqnarray}
\bm{A'} &=& \bm{A} + \varepsilon\bm{R} \\
\bm{f'} &=& \bm{f} + \varepsilon \bm{A'}^{-1} \bm{\eta},
\end{eqnarray}
where the $(3\times3)$ matrix $\bm{R}$ and the three-dimensional vector $\bm{\eta}$ are composed of uniform random numbers ranging from $-1$ to 1.
Parameter $\varepsilon$ is given to control the degree of lattice expansion, lattice distortion, and atomic displacements.

DFT calculations were performed for the structures in the datasets using the plane-wave-basis projector augmented wave method \cite{PAW1} within the Perdew--Burke--Ernzerhof exchange-correlation functional \cite{GGA:PBE96} as implemented in the \textsc{vasp} code \cite{VASP1,VASP2,PAW2}.
The cutoff energy was set to 300 eV.
The total energies converged to less than 10$^{-3}$ meV/supercell.
The atomic positions and lattice constants of the structure generators were optimized until the residual forces were less than 10$^{-2}$ eV/\AA.

\begin{figure}[tbp]
\includegraphics[clip,width=0.8\linewidth]{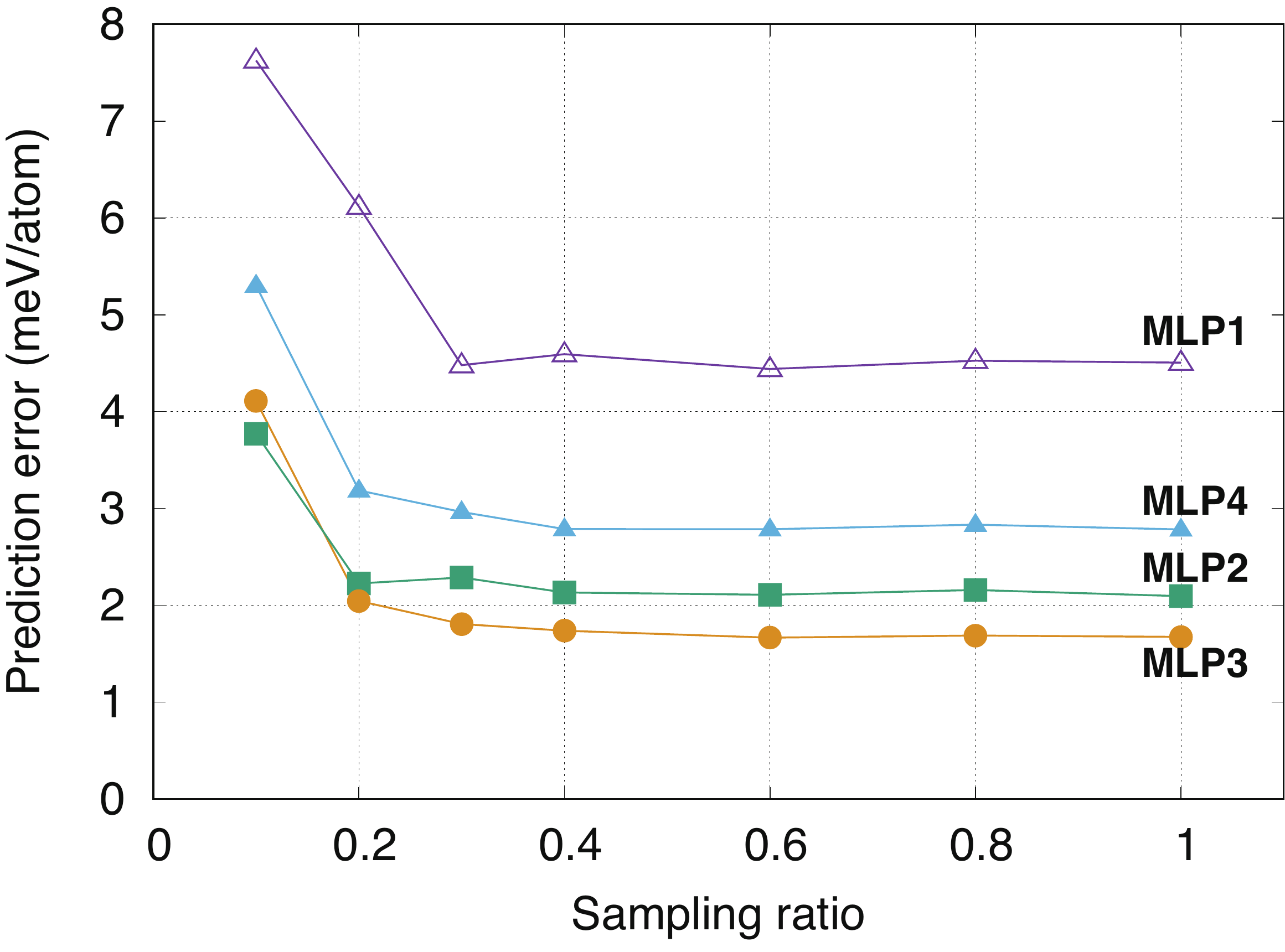}
\caption{
Training dataset size dependence of the prediction error for a test dataset in the binary Ti-Al system \cite{PhysRevB.102.174104}. 
A set of structures for each sampling ratio is randomly selected from the whole training dataset.
The number of structures in the whole training dataset is 41508, and the total number of entries is 5178510.
MLP1, MLP2, MLP3, and MLP4 consist of 7875, 27520, 61605, and 6940 regression coefficients, respectively.
}
\label{tutorial-2022:Fig-learning-curve}
\end{figure}

The accuracy of the polynomial MLP depends on the training dataset size.
Figure \ref{tutorial-2022:Fig-learning-curve} shows the training dataset size dependence of the prediction error for a test dataset in the binary Ti-Al system \cite{PhysRevB.102.174104}.
The size of the whole training dataset is 41508, although the prediction error converges well at a training dataset size of approximately 8000--12000.
Thus, the convergence of the prediction error in terms of the dataset size is important for developing robust polynomial MLPs.

It is worth emphasizing that the current datasets contain structures with much larger displacements than required to compute harmonic phonon force constants and structures with large lattice distortions.
As a result, the structures in the datasets show a wide range of the cohesive energy, although the range of the cohesive energy strongly depends on the system.
For example, the cohesive energy ranges from $-3.43$ eV/atom, corresponding to that of the most stable structure, to 5.00 eV/atom, corresponding to its given upper bound in elemental Al. 
The cohesive energy also ranges from $-4.55$ eV/atom to 5.00 eV/atom in elemental Si.
Contrary to this study, if the training and test datasets are composed of only structures with small displacements, the prediction errors of MLPs for a test dataset tend to be small.
Such an MLP is seemingly accurate because of the small prediction error. 
It may be accurate only in limited applications, whereas it should have low predictive power for many applications.

\subsection{Estimation of model coefficients}
\label{tutorial-2022:Sec-regression}

Coefficients of a polynomial MLP are estimated from all the total energies, forces, and stress tensors included in a training dataset.
Because the forces acting on atoms and the stress tensor are derived from the potential energy as linear models with its coefficients, the predictor matrix $\bm{X}$ and observation vector $\bm{y}$ can be simply written in a submatrix form as
\begin{equation}
\bm{X} =
\begin{bmatrix}
\bm{X}_{\rm energy} \\
\bm{X}_{\rm force} \\
\bm{X}_{\rm stress} \\
\end{bmatrix}
,\qquad \bm{y} =
\begin{bmatrix}
\bm{y}_{\rm energy} \\
\bm{y}_{\rm force} \\
\bm{y}_{\rm stress} \\
\end{bmatrix}.
\end{equation}
The predictor matrix $\bm{X}$ is composed of three submatrices, $\bm{X}_{\rm energy}$, $\bm{X}_{\rm force}$, and $\bm{X}_{\rm stress}$, which contain structural features and their polynomial contributions to the total energies, the forces acting on atoms, and the stress tensors of structures in the training dataset, respectively.
The elements of $\bm{X}_{\rm force}$ and $\bm{X}_{\rm stress}$ were derived in Ref. \onlinecite{PhysRevB.99.214108}.
The observation vector $\bm{y}$ also has three components, $\bm{y}_{\rm energy}$, $\bm{y}_{\rm force}$, and $\bm{y}_{\rm stress}$, which contain the total energy, the forces acting on atoms, and the stress tensors of structures in the training dataset, respectively, obtained from DFT calculations.
It is also possible to consider only the total energies and the forces acting on atoms.
In such a case, the predictor matrix and observation vector are represented by 
\begin{equation}
\bm{X} =
\begin{bmatrix}
\bm{X}_{\rm energy} \\
\bm{X}_{\rm force} \\
\end{bmatrix}
,\qquad \bm{y} =
\begin{bmatrix}
\bm{y}_{\rm energy} \\
\bm{y}_{\rm force} \\
\end{bmatrix}.
\end{equation}

Once the predictor matrix and the observation vector are evaluated, the coefficients $\bm{w}$ are estimated by linear ridge regression.
Linear ridge regression shrinks the regression coefficients by imposing a penalty and minimizes the penalized residual sum of squares expressed as
\begin{equation}
L(\bm{w}) = ||\bm{X}\bm{w} - \bm{y}||_2^2 + \lambda ||\bm{w}||_2^2,
\end{equation}
where $\lambda$ denotes the magnitude of the penalty.
This is referred to as $L_2$ regularization. 
The solution is represented as 
\begin{equation}
\bm{\hat w} = (\bm{X}^\top \bm{X} + \lambda \bm{I})^{-1} \bm{X}^\top \bm{y},
\end{equation}
where $\bm{I}$ denotes the unit matrix. 
The solution is easily obtained using linear algebra algorithms while avoiding the well-known multicollinearity problem occurring in the ordinary least-squares method.

Least absolute shrinkage and selection operator (Lasso) regression and elastic net regression, which combines the $L_1$ and $L_2$ penalties \cite{tibshirani1996regression,hastieelements}, are also applicable for estimating coefficients.
They deliver sparse solutions with a small number of nonzero coefficients.
Elastic net regression minimizes 
\begin{equation}
L(\bm{w}) = ||\bm{X}\bm{w} - \bm{y}||_2^2 + \alpha \lambda ||\bm{w}||_1 + \frac{(1-\alpha)}{2} \lambda ||\bm{w}||_2^2,
\label{lasso2:elastic_net_min_equation}
\end{equation}
where parameter $\alpha$ determines the mixing of the penalties, and the values of $\lambda$ and $\alpha$ simply control the accuracy and sparseness of the solution.
The minimization function with $\alpha = 1$ corresponds to that of the Lasso.
The Lasso and elastic net regression were introduced in the development of MLPs with linear models \cite{PhysRevB.90.024101,PhysRevB.92.054113}.

The computational implementation for estimating the model coefficients depends on the size of the predictor matrix.
When developing MLPs in elemental Al, the training dataset comprises 12242, 1373229, and 58866 entries for the energy, force, and stress tensor, respectively.
Therefore, the size of the predictor matrix $\bm{X}$ is $(1444337, n_{\rm coeff})$, where $n_{\rm coeff}$ denotes the number of coefficients of the potential energy model.
Since the number of coefficients ranges from seven to 23385 for elemental systems, the maximum memory required for allocating the predictor matrix is approximately 270 GB, available in standard workstations with a large amount of memory.
On the other hand, the required memory becomes huge in developing MLPs for alloy systems and several elemental systems that show complicated interatomic interactions.
For example, the MLP with the lowest RMS error for the binary Ti-Al system \cite{PhysRevB.102.174104} comprises 61605 coefficients.
In addition, an extensive training dataset of 41508 structures is used in the binary Ti-Al system, and the resultant total number of data entries is 5178510.
Therefore, the size of the predictor matrix becomes (5178510, 61605), which requires approximately 2.5 TB of memory.
If polynomial MLPs are developed for ternary and quaternary alloys, the situation is more severe, i.e., it is impossible to allocate the entire predictor matrix in a standard workstation.

When the predictor matrix is estimated to be large, a sequential implementation of linear ridge regression is practical, which is similar to an algorithm for updating model coefficients using appended data\cite{strang2019linear}.
In linear ridge regression, the evaluation of $\bm{X}^\top \bm{X}$ and $\bm{X}^\top \bm{y}$ is essential, whereas the predictor matrix $\bm{X}$ need not be evaluated.
Since the size of $\bm{X}^\top \bm{X}$, $(n_{\rm coeff}, n_{\rm coeff})$, is generally much smaller than the size of $\bm{X}$ in the current use of linear ridge regression, the following sequential implementation based on decomposing the training dataset into smaller batches significantly decreases the required memory.
The predictor matrix $\bm{X}$ and observation vector $\bm{y}$ can be represented with submatrix forms of
\begin{equation}
\bm{X} =
\begin{bmatrix}
\bm{X}_1 \\
\bm{X}_2 \\
\bm{X}_3 \\
\vdots \\
\end{bmatrix}
,\qquad \bm{y} =
\begin{bmatrix}
\bm{y}_1 \\
\bm{y}_2 \\
\bm{y}_3 \\
\vdots \\
\end{bmatrix},
\end{equation}
where $\bm{X}_i$ and $\bm{y}_i$ denote the predictor submatrix and the observation vector of the $i$th batch of the training dataset, respectively.
Using this decomposition, matrix $\bm{X}^\top \bm{X}$ and vector $\bm{X}^\top \bm{y}$ can be calculated without requiring the entire predictor matrix using the relationships 
\begin{eqnarray}
\bm{X}^\top \bm{X} &=& \left[ \bm{X}_1^\top \bm{X}_2^\top \bm{X}_3^\top \cdots \right]
\begin{bmatrix}
\bm{X}_1 \\
\bm{X}_2 \\
\bm{X}_3 \\
\vdots \\
\end{bmatrix} 
= \sum_i \bm{X}_i^\top \bm{X}_i 
\end{eqnarray}
and
\begin{eqnarray}
\bm{X}^\top \bm{y} &=& \left[ \bm{X}_1^\top \bm{X}_2^\top \bm{X}_3^\top \cdots \right]
\begin{bmatrix}
\bm{y}_1 \\
\bm{y}_2 \\
\bm{y}_3 \\
\vdots \\
\end{bmatrix} 
= \sum_i \bm{X}_i^\top \bm{y}_i .
\end{eqnarray}
The residual sum of squares (RSS) for the dataset is also computed without using predictor matrix $\bm{X}$ as
\begin{equation}
({\rm RSS}) = \bm{\hat w}^\top \left( \bm{X}^\top \bm{X} \right) \bm{\hat w} - 2 \bm{\hat w}^\top \left( \bm{X}^\top \bm{y} \right) + \bm{y}^\top \bm{y}.
\end{equation}

\begin{figure}[tbp]
\includegraphics[clip,width=0.8\linewidth]{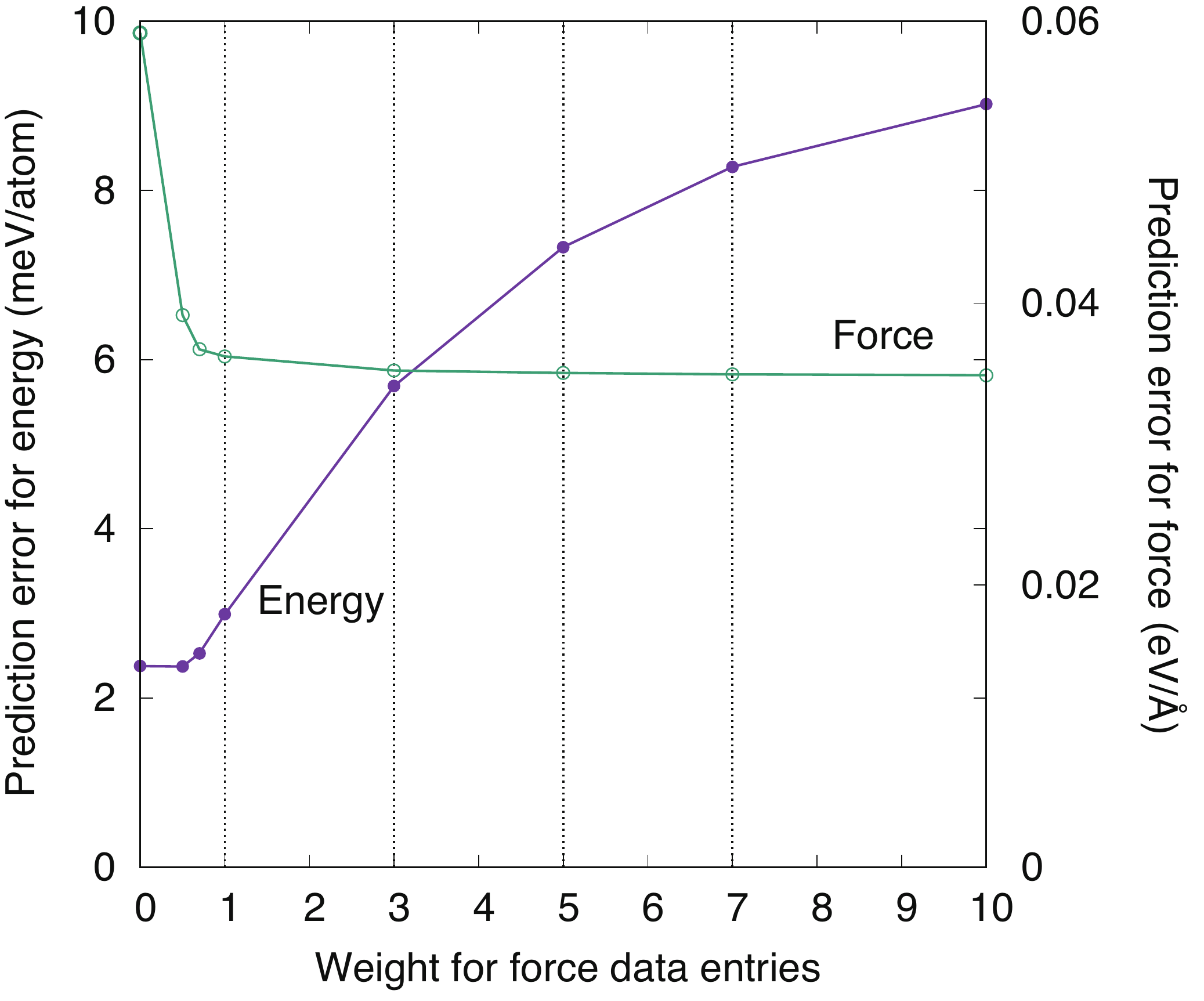}
\caption{
Dependence of the prediction errors for energy and force on the square root of the weight $\bm{W}^{1/2}$ for force data entries.
The prediction errors are estimated using a polynomial MLP model with 2030 coefficients for elemental Al.
Here, only the total energies and the forces acting on atoms are considered as the training dataset.
The units of eV/supercell and eV/\AA\ for the energy and force are respectively applied to the observations, and the weight for energy data entries is set to one.
}
\label{tutorial-2022:Fig-Al-weight}
\end{figure}

Also, it is indispensable to consider the weights that are given to data entries in the coefficient estimation.
Firstly, the units for the energy, force, and stress tensor in the observation vector are different. 
The units must be selected carefully because it is equivalent to the setting of weights in the regression.
Figure \ref{tutorial-2022:Fig-Al-weight} shows the dependence of the prediction errors for energy and force on the weight for force data entries.
Reasonable values of both prediction errors are obtained in the range of weights from 0.5 to 1.0 when the units of eV/supercell and eV/\AA\ are applied to the observations.
Here, the units of eV/supercell, eV/\AA\, and GPa are applied for the energy, forces, and stress tensor, respectively.
Moreover, if necessary, large weights should be imposed on important data entries such as those with low energies and small forces.
When weights are imposed on data entries, the solution of linear ridge regression is reformulated as
\begin{equation}
\bm{\hat w} = (\bm{X}^\top \bm{W} \bm{X} + \lambda \bm{I})^{-1} \bm{X}^\top \bm{W} \bm{y},
\end{equation}
where $\bm{W}$ denotes the diagonal matrix composed of weights.
This means that the weights modify matrix $\bm{X}$ and vector $\bm{y}$ to $\bm{X'} = \bm{W} ^{1/2} \bm{X}$ and $\bm{y'} = \bm{W}^{1/2} \bm{y}$, respectively, where $\bm{W}^{1/2}$ is the diagonal matrix whose $i$th diagonal element is the square root of the $i$th diagonal element of $\bm{W}$, $\sqrt{W_{ii}}$.

\subsection{Pareto optimality}
\label{tutorial-2022:Sec-method-pareto-optimality}
The accuracy and computational efficiency of the polynomial MLP strongly depend on the given input parameters.
They are 
(1) the cutoff radius, 
(2) the type of polynomial function $F$,
(3) the order of polynomial function $F$,
(4) the number of radial functions, 
and
(5) the truncation of the polynomial invariants, i.e., the maximum angular numbers of spherical harmonics $\{l_{\rm max}^{(2)}, l_{\rm max}^{(3)}, \cdots, l_{\rm max}^{(p_{\rm max})}\}$ and the maximum polynomial order of invariants $p_{\rm max}$.
Therefore, a systematic grid search is performed for each system to find their optimal values. 

As pointed out in Ref. \onlinecite{PhysRevB.99.214108}, the accuracy and computational efficiency are conflicting properties whose trade-off should be optimized.
However, there is no single optimal solution to this multiobjective optimization problem involving several conflicting objectives.
In such a case, Pareto-optimal points can be optimal solutions with different trade-offs between the accuracy and computational efficiency \cite{branke2008multiobjective}.
No points can simultaneously improve the accuracy and computational efficiency of the Pareto-optimal points.
Thus, the Pareto-optimal MLPs are obtained from the grid search for each system and distributed in the repository.

\section{Development of polynomial MLPs}
\label{tutorial-2022:Sec-mlp-repository}

\subsection{Pareto-optimal MLPs}

\begin{figure}[tbp]
\includegraphics[clip,width=0.8\linewidth]{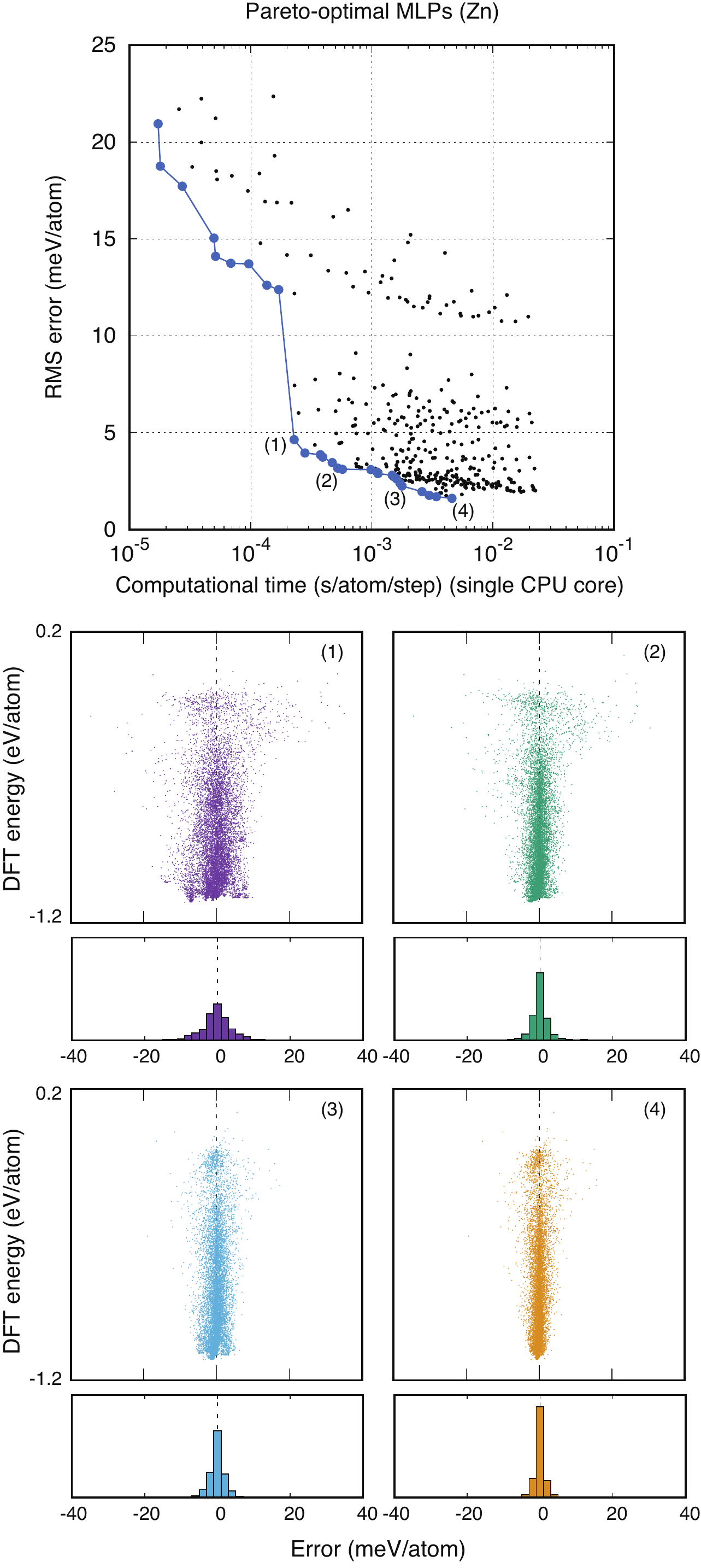}
\caption{
Distribution of MLPs for elemental Zn obtained from a grid search to find the optimal model parameters.
The elapsed time for a single point calculation is estimated using a single core of Intel\textregistered\ Xeon\textregistered\ E5-2695 v4 (2.10 GHz).
The blue closed circles show the Pareto-optimal points of the distribution obtained using a non-dominated sorting algorithm.
Distributions of the errors in the training and test datasets for selected MLPs and their histograms with a bin size of 2 meV/atom are also shown.
}
\label{tutorial-2022:Fig-Zn-pareto-error}
\end{figure}

\begin{table}[tbp]
\begin{ruledtabular}
\caption{
Model parameters of the Pareto-optimal MLPs for elemental Zn selected in Fig. \ref{tutorial-2022:Fig-Zn-pareto-error}.
}
\label{tutorial-2022:Zn-parameter}
\begin{tabular}{lcccc}
& (1) & (2) & (3) & (4) \\
\hline
Number of coefficients & 945 & 2030 & 10010 & 19502 \\
Cutoff radius & 6.0 & 7.0 & 7.0 & 7.0 \\
Number of radial functions & 7 & 10 & 7 & 7 \\
Polynomial order (function $F$) & 2 & 2 & 2 & 2 \\
Polynomial order (invariants)   & 2 & 3 & 3 & 3 \\
$\set{l_{\rm max}^{(2)}, l_{\rm max}^{(3)},\cdots}$ & [4] & [4,4] & [4,4] & [12,4] 
\end{tabular}
\end{ruledtabular}
\end{table}

\begin{figure}[tbp]
\includegraphics[clip,width=\linewidth]{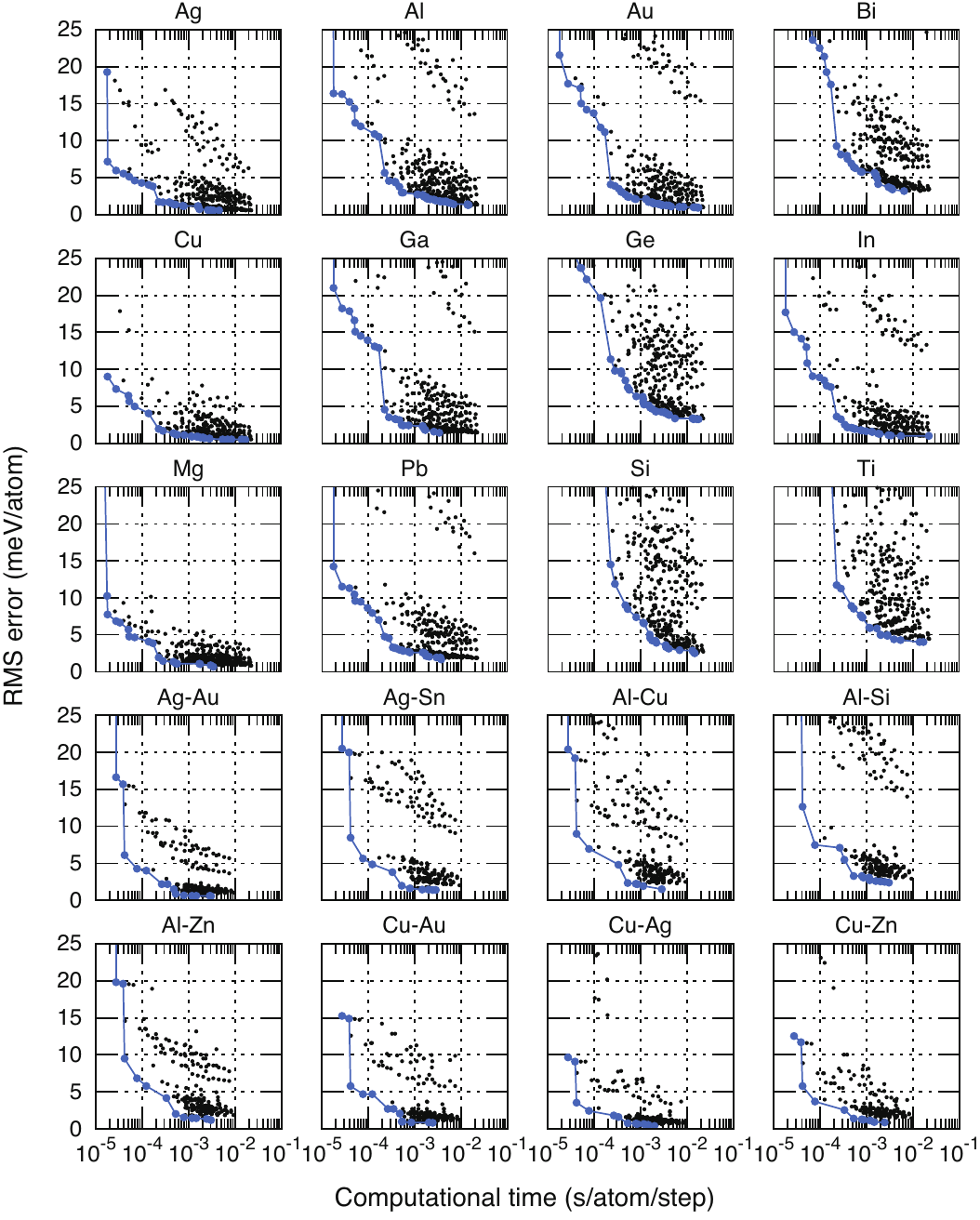}
\caption{
Distributions of MLPs for elemental Ag, Al, Au, Bi, Cu, Ga, Ge, In, Mg, Pb, Si and Ti and for the binary Ag-Au, Ag-Sn, Al-Cu, Al-Si, Al-Zn, Cu-Au, Cu-Ag, and Cu-Zn alloys obtained from grid searches for the model parameters.
The blue closed circles show the Pareto-optimal points of the distributions.
The distributions for the other systems can be found in the repository\cite{MachineLearningPotentialRepository}.
}
\label{tutorial-2022:Fig-pareto-elements-alloy}
\end{figure}

Figure \ref{tutorial-2022:Fig-Zn-pareto-error} shows the prediction error and computational efficiency of the Pareto-optimal MLPs for elemental Zn obtained by a grid search of model parameters.
Figure \ref{tutorial-2022:Fig-pareto-elements-alloy} shows the distributions of MLPs and Pareto-optimal MLPs for elemental Ag, Al, Au, Bi, Cu, Ga, Ge, In, Mg, Pb, Si, and Ti and for the binary Ag-Au, Ag-Sn, Al-Cu, Al-Si, Al-Zn, Cu-Au, Cu-Ag, and Cu-Zn alloys.
The prediction error is estimated using the root mean square (RMS) error of the energy for the test dataset.
The computational efficiency is estimated using the elapsed time to compute the energy, forces, and stress tensors of a structure with 284 atoms.
The elapsed time is normalized by the number of atoms because it is proportional to the number of atoms, as shown later.
As can be seen in Figs. \ref{tutorial-2022:Fig-Zn-pareto-error} and \ref{tutorial-2022:Fig-pareto-elements-alloy}, the accuracy and computational efficiency of MLPs are conflicting properties; hence, the Pareto-optimal MLPs can be candidates for use in subsequent simulations of interest.
Figure \ref{tutorial-2022:Fig-Zn-pareto-error} also shows the distributions of errors in the training and test datasets for the four Pareto-optimal MLPs of elemental Zn.
Table \ref{tutorial-2022:Zn-parameter} lists the values of the model parameters of the four MLPs.
As the RMS error decreases, the variance of the distribution becomes small.

Note that the magnitude of the RMS error strongly depends on the intrinsic properties of systems, such as the bulk modulus, the elastic constants, the magnitude of phonon frequencies, and the energy distribution of structures in the dataset.
Therefore, it is preferable to use RMS errors normalized by such properties when evaluating the contribution of the RMS error to the prediction of properties and comparing the accuracy of MLPs in different systems. 
For example, Takahashi et al. normalized the RMS error by the standard deviation of the energies for datasets to compare the accuracy of MLPs for different elemental systems \cite{doi:10.1063/1.5027283}.
Consequently, in some elemental systems, the normalized RMS error is substantial despite the small RMS error, which means that the inconsistency of the physical properties between the MLP and the DFT calculation should be regarded as significant despite the RMS error being small.

\begin{figure}[tbp]
\includegraphics[clip,width=0.8\linewidth]{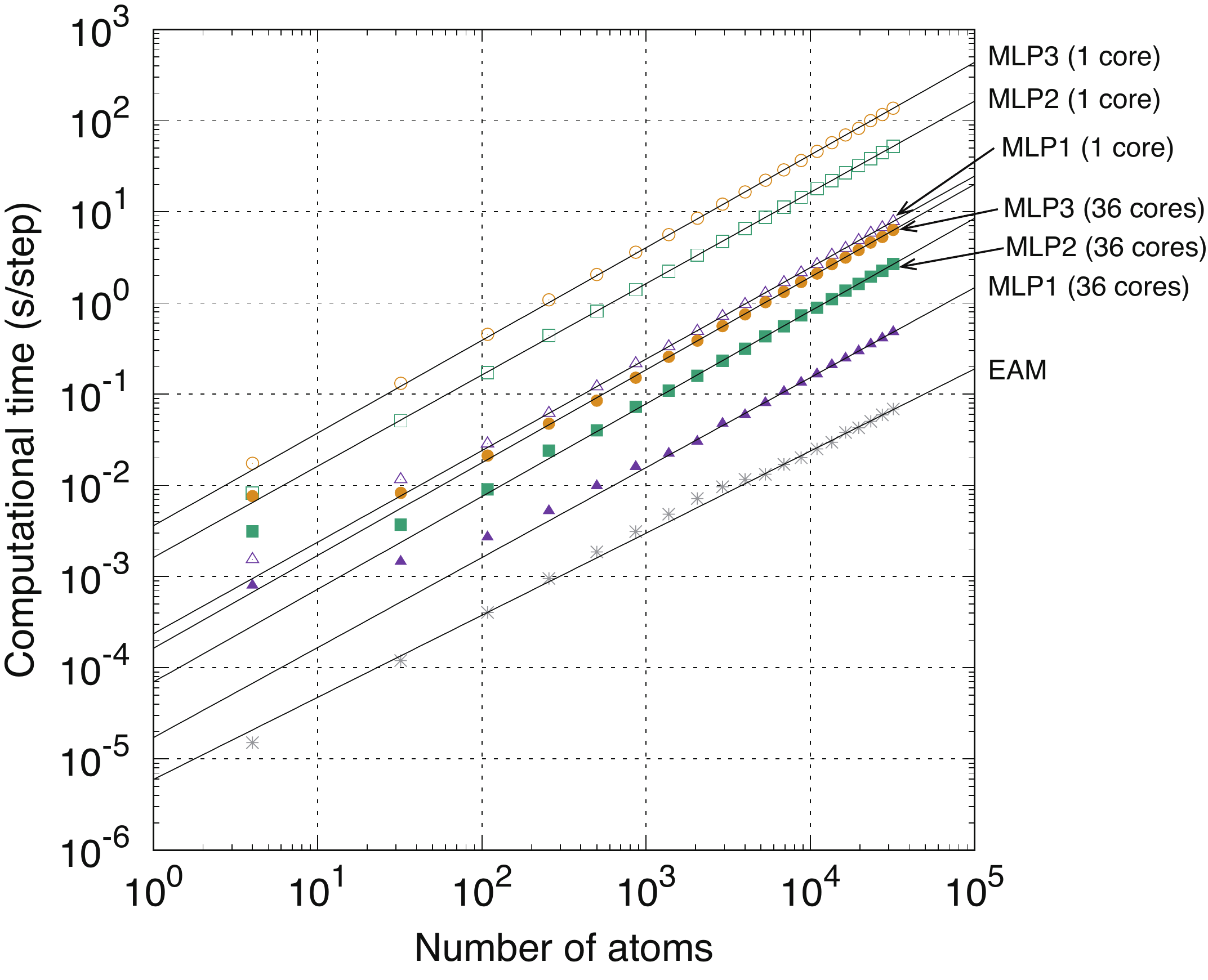}
\caption{
Dependence of the computational time required for a single point calculation on the number of atoms.
The elapsed time is measured using a single core and 36 cores of Intel\textregistered\ Xeon\textregistered\ E5-2695 v4 (2.10 GHz).
An implementation of the polynomial MLP in \textsc{lammps} (\textsc{lammps-polymlp}), which is compatible with \textsc{openmp} support \cite{LammpsPolyMLP}, is used to measure the elapsed time.
The computations using 36 cores are approximately 25 times as fast as those using a single core.
}
\label{tutorial-2022:Fig-Ti-Al-time-scale}
\end{figure}

Figure \ref{tutorial-2022:Fig-Ti-Al-time-scale} shows the elapsed times of single point calculations for structures with up to 32000 atoms using the EAM potential \cite{PhysRevB.68.024102} and the three Pareto-optimal MLPs for the binary Ti-Al system \cite{PhysRevB.102.174104}.
Structures were made by the expansion of the fcc conventional unit cell with a lattice constant of 4 \AA.
As can be seen in Fig. \ref{tutorial-2022:Fig-Ti-Al-time-scale}, linear scaling with respect to the number of atoms is achieved in all the MLPs.
Although the performance for the three MLPs only in the binary Ti-Al system is shown here, the other MLPs also exhibit linear scaling with respect to the number of atoms.
Therefore, the computational time required for a calculation of $n_{\rm step}$ steps for a structure with $n_{\rm atom}$ atoms can be roughly estimated as $ t \times n_{\rm atom} \times n_{\rm step}$, where $t$ is the elapsed time per atom for a single point calculation listed in the repository.

At this time, the Pareto-optimal polynomial MLPs are available in the repository website for elemental Ag, Al, As, Au, Ba, Be, Bi, Ca, Cd, Cr, Cs, Cu, Ga, Ge, Hf, Hg, In, Ir, K, La, Li, Mg, Mo, Na, Nb, Os, P, Pb, Pd, Pt, Rb, Re, Rh, Ru, Sb, Sc, Si, Sn, Sr, Ta, Te, Ti, Tl, V, W, Y, Zn and Zr and binary Ag-Au, Ag-Ba, Ag-In, Ag-La, Ag-Pb, Ag-Sn, Al-Ag, Al-Au, Al-Ba, Al-Ca, Al-Cu, Al-Ga, Al-Ge, Al-In, Al-K, Al-La, Al-Nb, Al-Pb, Al-Pt, Al-Sc, Al-Si, Al-Sn, Al-Sr, Al-V, Al-Y, Al-Zn, Al-Zr, Be-Al, Be-Ge, Be-Si, Be-Sn, Ca-Ag, Ca-Au, Ca-Cu, Ca-Ga, Ca-Ge, Ca-In, Ca-Pb, Ca-Sc, Ca-Sn, Ca-Ti, Ca-Zn, Cu-Ag, Cu-Au, Cu-Zn, Ga-Ag, Ga-Sn, Ge-In, Ge-Pb, Ge-Sn, Ge-Sr, K-Ge, K-Pb, K-Sn, K-Zn, Li-Ag, Li-Al, Li-Au, Li-Ba, Li-Be, Li-Bi, Li-Ca, Li-Cu, Li-Ga, Li-Ge, Li-In, Li-Mg, Li-Pb, Li-Si, Li-Sn, Li-Sr, Li-Ti, Li-Zn, Mg-Ag, Mg-Al, Mg-Au, Mg-Ca, Mg-Cu, Mg-Ga, Mg-Ge, Mg-In, Mg-K, Mg-Sc, Mg-Si, Mg-Sn, Mg-Ti, Mg-Y, Na-Al, Na-Ga, Na-Ge, Na-In, Na-Mg, Na-Pb, Na-Si, Na-Sn, Pd-Pt, Si-Ag, Si-Au, Si-Cu, Si-Ga, Si-Ge, Si-In, Si-Pb, Si-Sn, Si-Zn, Sn-Pb, Sr-Pb, Ti-Al, Zn-Ag, Zn-Ga, Zn-Ge, Zn-In, and Zn-Sn alloy systems.
For each system, the behavior of the relationship between the prediction error and the computational efficiency can be found in the repository.
For each Pareto-optimal MLP, its estimated accuracy, its computational efficiency, the input model parameters used for developing the MLP, and the distributions of the predicted energies can also be found in the repository.

\subsection{Selection from Pareto-optimal MLPs}
An appropriate MLP must be chosen from the Pareto-optimal ones to perform an atomistic simulation according to its target system and purpose.
The current MLPs often exhibit high predictive power for the properties shown in the following section.
However, it should be noted that the current MLPs are not necessarily accurate for some properties and for structures far from those in the training dataset.
Therefore, it is necessary to examine the predictive power for properties relevant to the target properties and their convergence behaviors in terms of the computational cost using the whole set of Pareto-optimal MLPs.
If there are no Pareto-optimal MLPs with high predictive power for the properties of interest, additional DFT calculations for their related structures and the re-estimation of model coefficients for the MLPs are desirable.

\begin{figure}[tbp]
\includegraphics[clip,width=\linewidth]{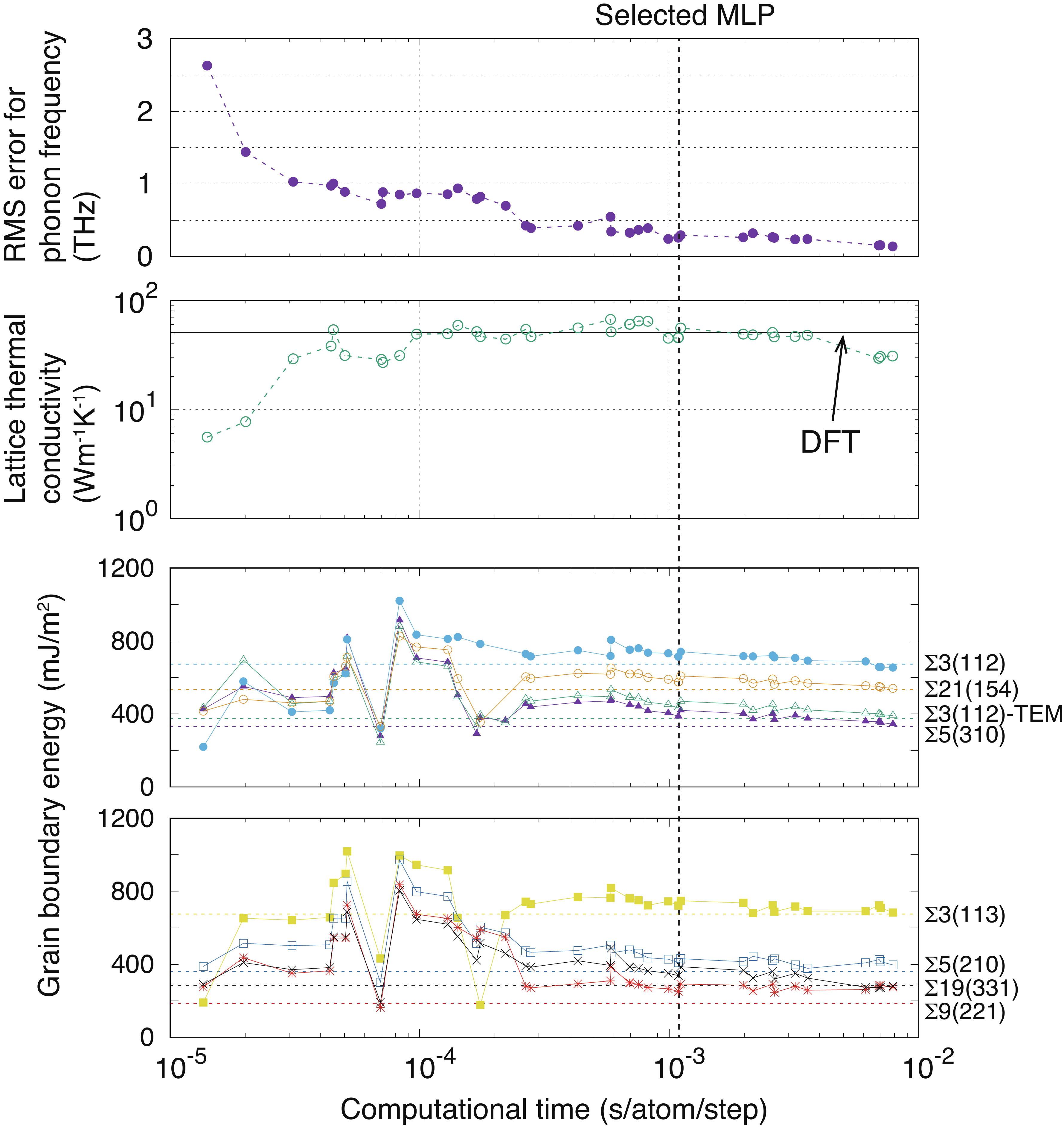}
\caption{
Convergence behaviors of the RMS error for the phonon frequency of the diamond structure, the lattice thermal conductivity of the diamond structure, and the grain boundary energies with respect to the computational time of MLP in silicon \cite{FUJII2022111137}.
In the lower panels, the broken horizontal lines indicate the grain boundary energies computed by the DFT calculation.
The phonon properties and lattice thermal conductivity are calculated by the finite displacement method using \textsc{phonopy} \cite{PhonopyArticle} and \textsc{phono3py} \cite{PhysRevB.91.094306}, respectively.
The lattice thermal conductivity is calculated by solving the Boltzmann transport equation with the single-mode relaxation-time approximation.
}
\label{tutorial-2022:Fig-Si-convergence}
\end{figure}

Fujii and Seko investigated the grain boundary structures, grain boundary phonon properties, and lattice thermal conduction at grain boundaries using large-scale perturbed molecular dynamics and phonon wave-packet simulations with a Pareto-optimal polynomial MLP in silicon \cite{FUJII2022111137}.
Therefore, a reasonable MLP was chosen from the Pareto-optimal MLPs in the repository by examining the convergence behaviors of the cohesive energies, the RMS error for the phonon frequency of the diamond structure, the lattice thermal conductivity of the diamond structure, and grain boundary energies.
Figure \ref{tutorial-2022:Fig-Si-convergence} shows the convergence behaviors of the RMS error for the phonon frequency of the diamond structure, the lattice thermal conductivity of the diamond structure, and grain boundary energies in silicon.
The RMS error for the phonon frequency approaches zero, and the lattice thermal conductivity and the grain boundary energies converge for the computational time of the Pareto-optimal MLP.
Also, Nishiyama et al. demonstrated that the MLP with the lowest computational time among the MLPs showing convergence for the grain boundary energy could accurately predict the grain boundary energies of larger models \cite{PhysRevMaterials.4.123607}.

Thus, the convergence behaviors of the Pareto-optimal MLPs can play an essential role in finding a reasonable MLP.
In the repository website, the convergence behaviors of the cohesive energy and the equilibrium volume for selected structures and the dependence of other properties such as elastic constants on the Pareto-optimal MLP can be easily accessed \cite{MachineLearningPotentialRepository}.
They should help users select an MLP from the Pareto-optimal MLPs.

Based on the convergence behaviors, it is desirable to select MLPs showing high computational cost performance.
In other words, they exhibit high computational efficiency without significantly increasing the prediction error.
A procedure to determine such an MLP is to evaluate the convex hull of the distribution of MLPs.
The convex hull provides a set of MLPs minimizing $ at + \Delta E$ for some positive values of parameter $a$, where $\Delta E$ and $t$ respectively denote the RMS error for the energy and the computational time per atom for a single point calculation.
Since $ at + \Delta E$ is regarded as a score for measuring the computational cost performance, the MLPs on the convex hull showing the converged values of properties can be candidates for use in the subsequent simulations.

\section{Predictive power of polynomial MLPs}
\label{tutorial-2022:Sec-predictive-power}

This section shows the predictive power of only a small part of the polynomial MLPs in the repository, whereas predictions of properties using the other MLPs can be found in the repository website \cite{MachineLearningPotentialRepository}.
Although this paper mainly reviews the predictive power for properties that were reported in elemental Ag, Al, Au, Cu, Pd, Pt, and Si and the binary Ti-Al alloy \cite{PhysRevB.99.214108,PhysRevB.102.174104,PhysRevMaterials.4.123607,FUJII2022111137}, the predictive power of the other MLPs will be reported elsewhere in the near future.


\subsection{Cohesive energy and volume}

\begin{figure}[tbp]
\includegraphics[clip,width=0.9\linewidth]{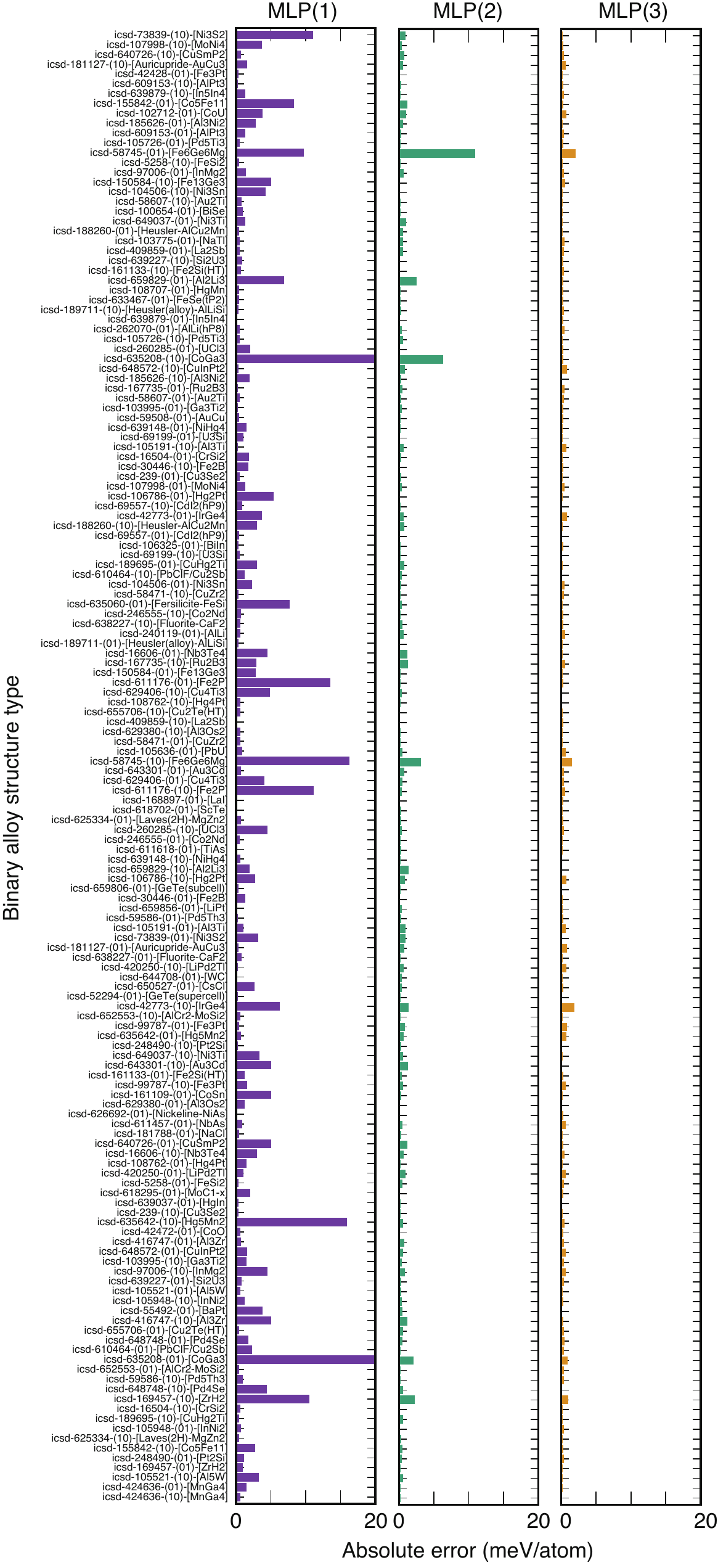}
\caption{
Absolute prediction errors of the cohesive energy for the 150 alloy prototype structures in the binary Ag-Au alloy. 
They correspond to the structure generators used for generating the training and test datasets.
Identifiers (01) and (10) in the labels of structure types respectively indicate the original alloy configuration and a configuration where the elements in the original one are swapped.
}
\label{tutorial-2022:Fig-Ag-Au-icsd-bar}
\end{figure}

Figure \ref{tutorial-2022:Fig-Ag-Au-icsd-bar} shows the absolute prediction errors of the cohesive energy for various alloy structures in the binary Ag-Au alloy.
MLP(1), MLP(2), and MLP(3) are composed of 3255, 27520, and 63855 coefficients and exhibit RMS errors of 4.27, 0.92, and 0.58 meV/atom, respectively.
MLP(1) shows large errors for a small number of structures and relatively small errors for the other structures.
The use of complex models significantly improves the accuracy for the whole range of structures.
Therefore, the convergence of the cohesive energy for many structures with respect to the computational cost of the MLP is often faster than that of the RMS error.
Similar behaviors of the error distribution and accuracy improvement can also be recognized in the other elemental and alloy systems.

\begin{figure}[tbp]
\includegraphics[clip,width=0.9\linewidth]{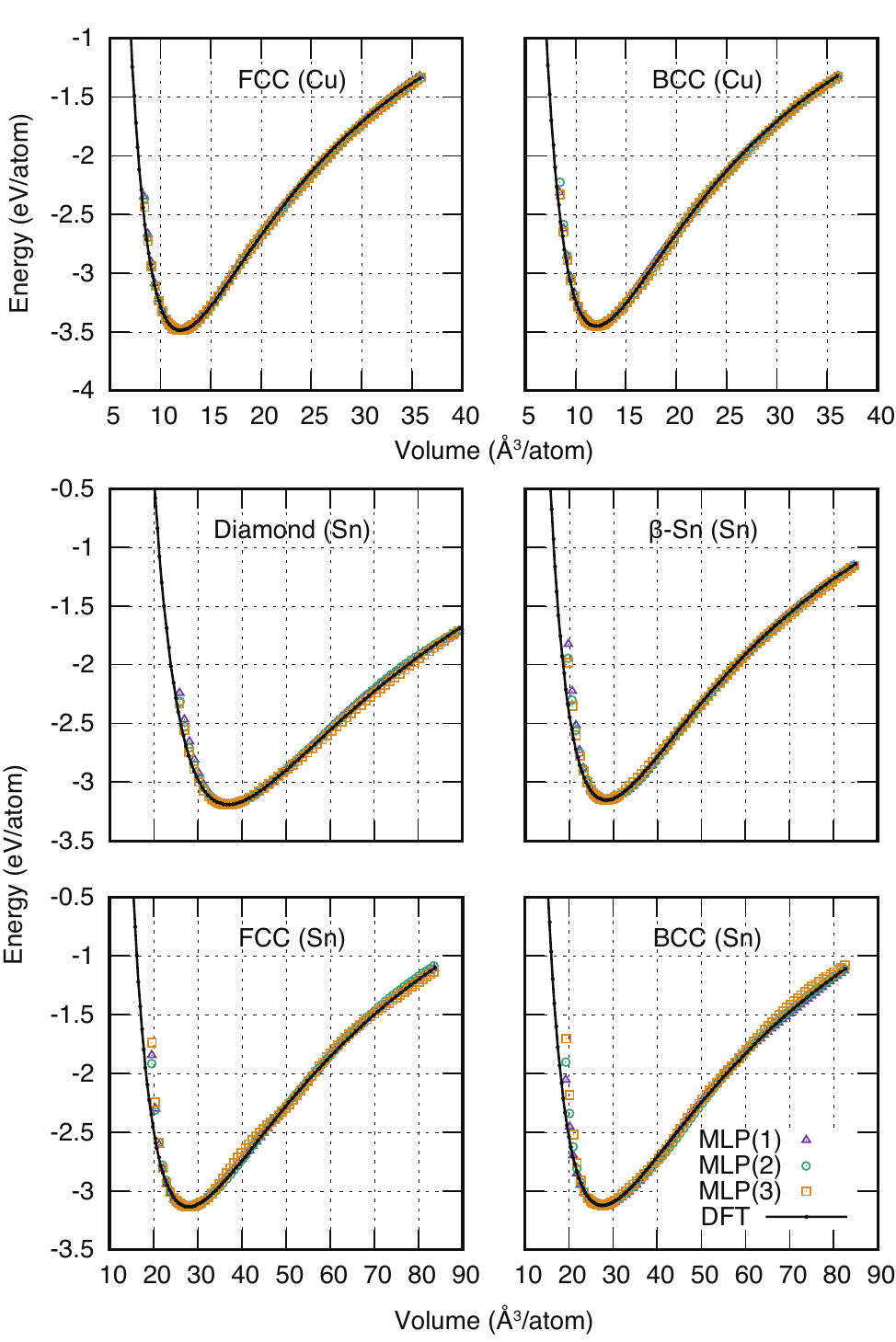}
\caption{
Energy--volume curves of the fcc and bcc structures predicted using selected MLPs in elemental Cu and those of the diamond, $\beta$-Sn, fcc, and bcc structures in elemental Sn, compared with those obtained by the DFT calculation.
For elemental Cu, MLP(1), MLP(2), and MLP(3) are composed of 945, 5830, and 23385 coefficients and exhibit RMS errors of 1.9, 0.7, and 0.5 meV/atom, respectively.
For elemental Sn, MLP(1), MLP(2), and MLP(3) are composed of 5420, 5700, and 23385 coefficients and show RMS errors of 3.1, 2.7, and 2.1 meV/atom, respectively.
}
\label{tutorial-2022:Fig-ev}
\end{figure}

Figure \ref{tutorial-2022:Fig-ev} shows the energy--volume curves of the face-centered cubic (fcc) and body-centered cubic (bcc) structures in elemental Cu and those of the diamond, $\beta$-Sn, fcc, and bcc structures in elemental Sn.
In the other elemental and binary alloy systems, predicted equations of states for various structures are shown for each Pareto-optimal MLP in the repository.
As can be seen in Fig. \ref{tutorial-2022:Fig-ev}, the equations of states predicted using the MLPs are almost consistent with those obtained from the DFT calculation.

\begin{figure*}[tbp]
\includegraphics[clip,width=\linewidth]{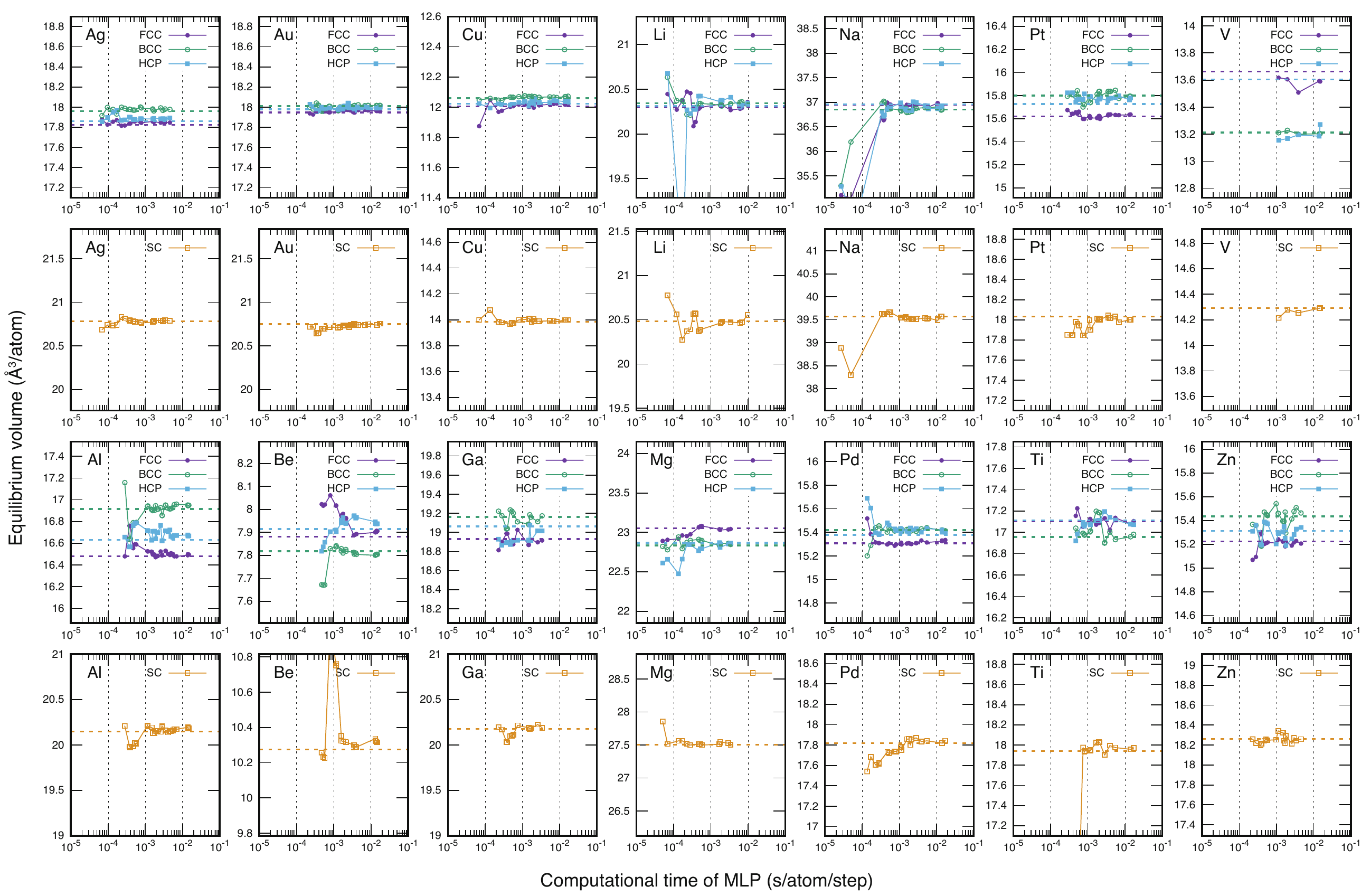}
\caption{
Dependence of the equilibrium volume on the selection of the MLP for the fcc, bcc, hcp, and sc structures in elemental Ag, Al, Au, Be, Cu, Ga, Li, Mg, Na, Pd, Pt, Ti, V, and Zn.
The broken lines show the equilibrium volumes obtained by the DFT calculation.
The ranges of the equilibrium volumes corresponding to the vertical axes are set to 10 \% of the equilibrium volumes for the fcc and sc structures.
}
\label{tutorial-2022:Fig-volumes}
\end{figure*}

Figure \ref{tutorial-2022:Fig-volumes} shows the convergence behaviors of the equilibrium volumes for the fcc, bcc, hexagonal close-packed (hcp), and simple cubic (sc) structures in elemental Ag, Al, Au, Be, Cu, Ga, Li, Mg, Na, Pd, Pt, Ti, V, and Zn.
As the model complexity increases, the equilibrium volume predicted using the MLP approaches that obtained from the DFT calculation in most of the cases and also in the other elemental systems, as found in the repository. 
Moreover, the final structure of the local geometry optimization is usually independent of the selection of the MLP for many structure types, which indicates that the potential energy surfaces around the equilibrium structure predicted using the Pareto-optimal MLPs are similar.
The structure type of the final structure is generally but not always the same as that of the initial structure.
On the other hand, the structure type of the final structure sometimes depends on the selection of the MLP.
In such a case, care is required in selecting MLPs because the dependence of the equilibrium volume on the computational cost of the MLP is not smooth.

In Fig. \ref{tutorial-2022:Fig-volumes}, a different structure with space group ($Fmmm$) is obtained by a local geometry optimization from the hcp structure in elemental V, which results in a significant deviation between the equilibrium volumes computed from the MLPs and the DFT calculation. 
In this case, the deviation is ascribed to a different treatment of symmetry constraints in local optimization algorithms implemented in \textsc{lammps} and \textsc{vasp}, which are used for the MLP and DFT calculations, respectively.

\subsection{Formation energy}

\begin{figure}[tbp]
\includegraphics[clip,width=\linewidth]{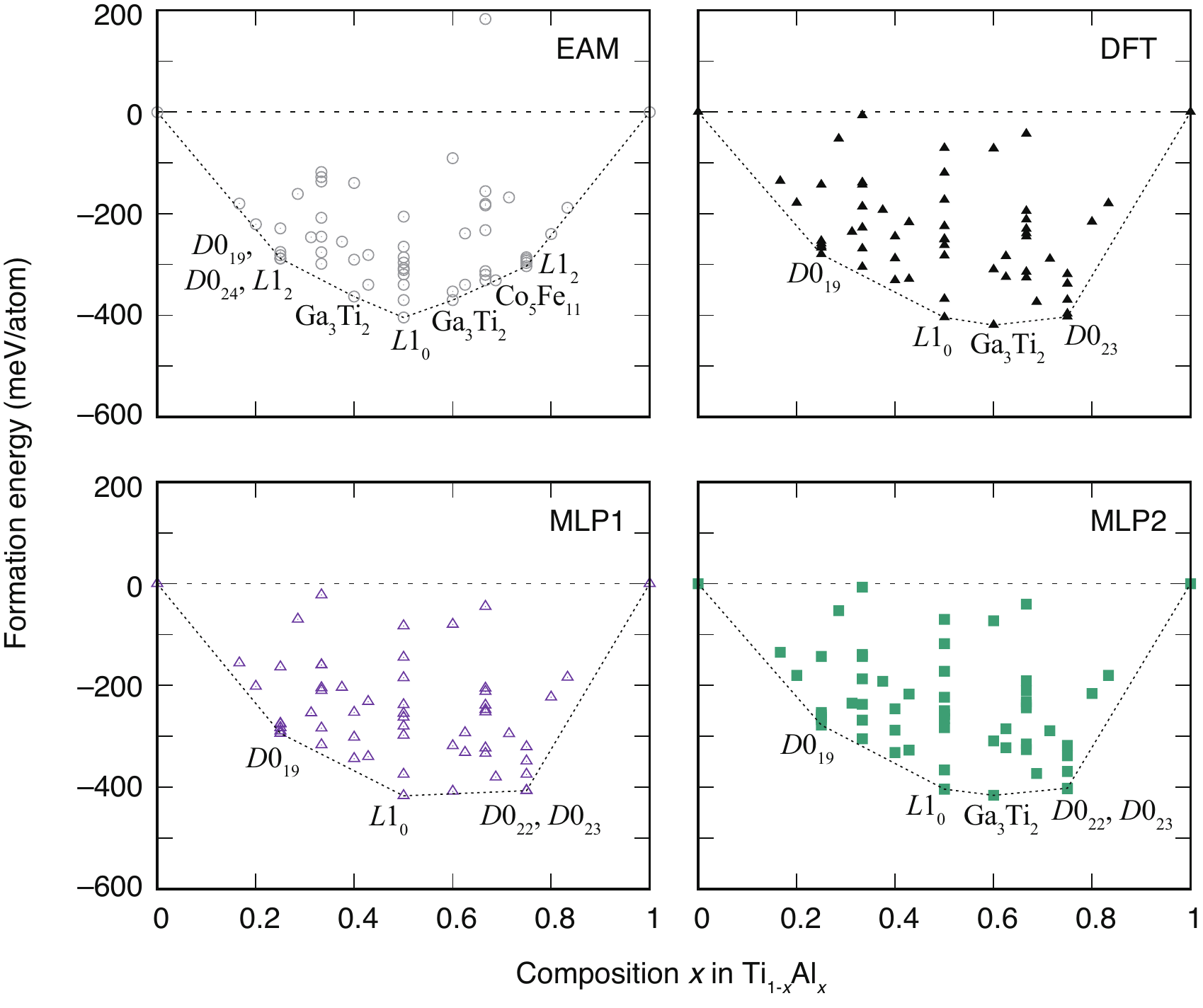}
\caption{
Formation energies of selected ordered structures in the binary Ti-Al alloy system computed by the EAM potential \cite{PhysRevB.68.024102}, the DFT calculation, and two Pareto-optimal MLPs \cite{PhysRevB.102.174104}.
The convex hull obtained from the formation energies of selected ordered structures is also shown.
MLP1 and MLP2 are composed of 7875 and 27520 coefficients and show RMS errors of 4.5 and 2.1 meV/atom, respectively.
}
\label{tutorial-2022:Fig-Ti-Al-formation-energy}
\end{figure}

The high predictive power for the alloy formation energy enables us to predict the phase stability between ordered structures and accomplish accurate crystal structure searches in alloy systems.
On the other hand, the formation energy of a given ordered structure is more challenging to predict accurately than its cohesive energy because high predictive power is required for not only the ordered structure but also the reference structures, e.g., hcp Ti and fcc Al in the Ti-Al system.
Furthermore, the local geometry relaxation for the ordered structure is crucial for evaluating the formation energy.
Therefore, MLPs are required to derive an accurate potential energy surface around the initial and equilibrium structures.

Figure \ref{tutorial-2022:Fig-Ti-Al-formation-energy} shows the formation energies of selected ordered structures predicted using the EAM potential and two Pareto-optimal MLPs compared with those predicted by DFT calculation.
The RMS errors of the EAM potential, MLP1, and MLP2 for the formation energy are 68.0, 13.8, and 2.0 meV/atom, although MLP1 and MLP2 show RMS errors of 4.5 and 2.1 meV/atom computed from the test dataset, respectively.
This reveals that MLP2 has high predictive power for the formation energy in a wide range of structures.
On the other hand, the RMS error of MLP1 for the formation energy is greater than the RMS error for the energy computed from the test dataset.
This is ascribed to the systematic deviation of the formation energy for the overall ordered structures.
The formation energies of most of the ordered structures predicted using MLP1 are approximately 10 meV/atom lower than those predicted by DFT calculation, which originates from the fact that the prediction error of MLP1 for hcp-Ti is large ($+$27.4 meV/atom).

\subsection{Elastic constants and phonon-related properties}

\begin{figure}[tbp]
\includegraphics[clip,width=\linewidth]{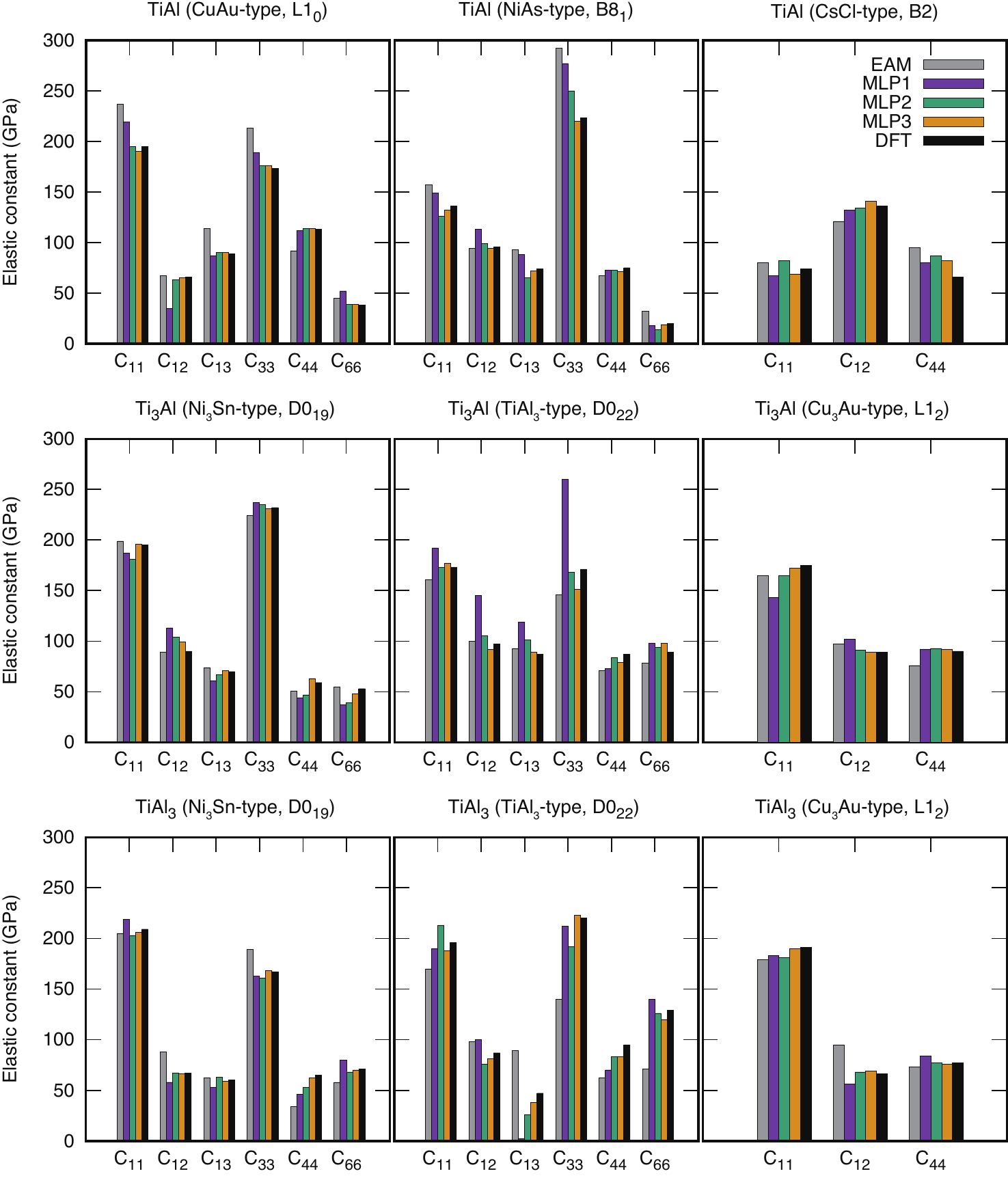}
\caption{
Elastic constants for selected ordered structures in the Ti-Al system calculated using the EAM potential\cite{PhysRevB.68.024102}, the MLPs\cite{PhysRevB.102.174104}, and the DFT calculation.
MLP1, MLP2, and MLP3 show RMS errors (force) of 0.138, 0.074, and 0.066 eV/\AA, respectively.
}
\label{tutorial-2022:Fig-Ti-Al-elastic-const}
\end{figure}

\begin{figure}[tbp]
\includegraphics[clip,width=\linewidth]{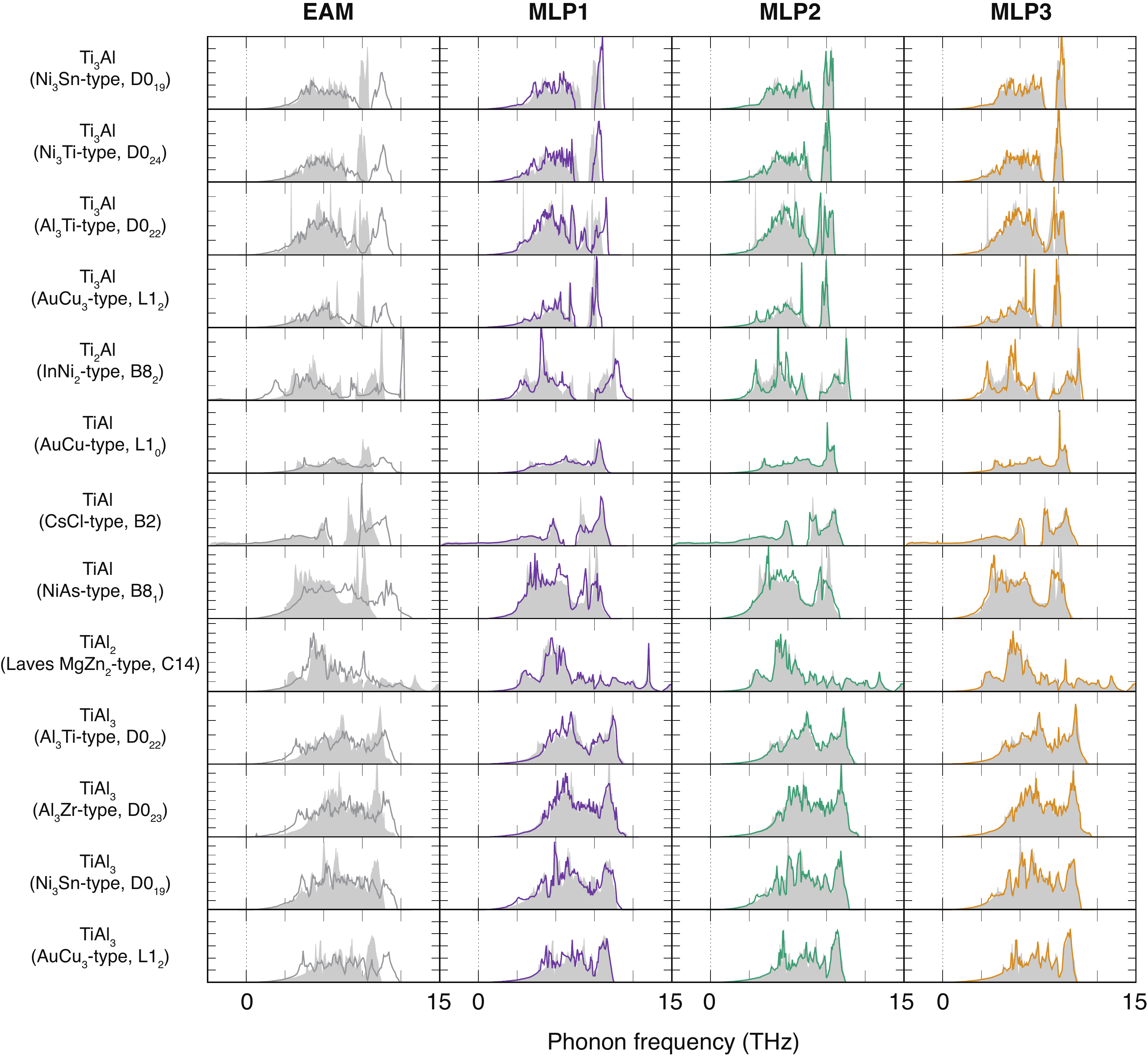}
\caption{
Phonon DOS for selected compounds in the Ti-Al binary system predicted using the EAM potential\cite{PhysRevB.68.024102} and the MLPs\cite{PhysRevB.102.174104}.
The shaded region indicates the phonon DOS computed by DFT calculation.
MLP1, MLP2, and MLP3 are composed of 7875, 27520, and 61605 coefficients and show RMS errors (force) of 0.138, 0.074, and 0.066 eV/\AA, respectively.
}
\label{tutorial-2022:Fig-Ti-Al-phonon-dos}
\end{figure}

The predictive power of the MLPs for the elastic constants, phonon properties, and thermal expansion is examined.
The predicted elastic constants for nine ordered structures in the Ti-Al system\cite{PhysRevB.102.174104} are shown in Fig. \ref{tutorial-2022:Fig-Ti-Al-elastic-const}. 
The EAM potential and the MLP with large RMS errors fail to predict the elastic constants accurately in a few structures, whereas the elastic constants predicted using the MLPs with low RMS errors are almost the same as those obtained by DFT calculation.
The phonon properties and thermal expansion are calculated using a finite displacement method implemented in the \textsc{phonopy} code \cite{PhonopyArticle}.
Figure \ref{tutorial-2022:Fig-Ti-Al-phonon-dos} shows the phonon density of states (DOS) for 13 structures predicted using the EAM potential and three MLPs.
The EAM potential predicts the phonon DOS well in the low-frequency region for many structures, whereas the deviation from the DFT phonon DOS is large in the high-frequency region.
Conversely, the phonon DOS predicted using the MLPs and those predicted by DFT calculation overlap for all the structures, particularly those predicted using the two accurate MLPs.

\begin{figure}[tbp]
\includegraphics[clip,width=\linewidth]{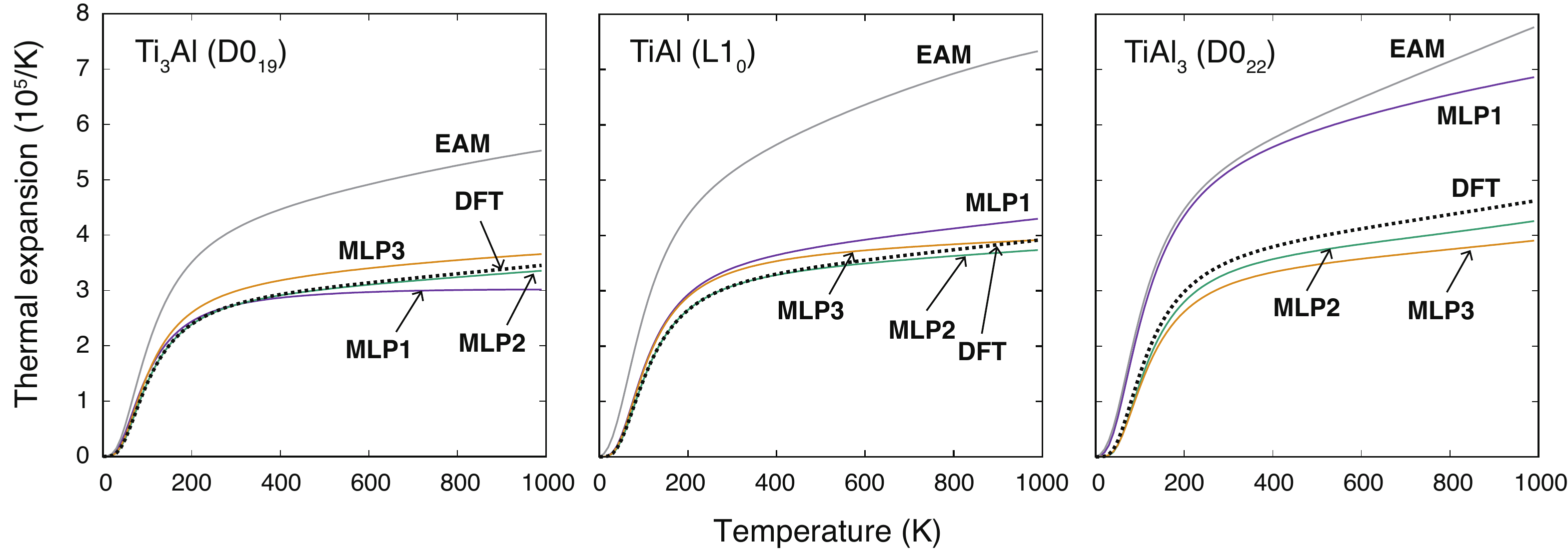}
\caption{
Temperature dependence of the thermal expansion predicted using the EAM potential\cite{PhysRevB.68.024102} and the MLPs for the three ordered structures in the Ti-Al system\cite{PhysRevB.102.174104}.
The broken black lines show the thermal expansion computed by DFT calculation.
MLP1, MLP2, and MLP3 show RMS errors for the energy of 4.5, 2.1, and 1.7 eV/atom and RMS errors for the force of 0.138, 0.074, and 0.066 eV/\AA, respectively.
}
\label{tutorial-2022:Fig-Ti-Al-thermal-expansion}
\end{figure}

The thermal expansion is more challenging to predict accurately than the phonon DOS and the phonon dispersion curves.
Figure \ref{tutorial-2022:Fig-Ti-Al-thermal-expansion} shows the temperature dependence of the thermal expansion, calculated using a quasi-harmonic approximation, in Ti$_3$Al ($D0_{19}$), TiAl ($L1_0$), and TiAl$_3$ ($D0_{22}$), which are experimentally observed in the Ti-Al binary system.
As can be seen in Fig. \ref{tutorial-2022:Fig-Ti-Al-thermal-expansion}, the thermal expansion of the EAM potential differs from that of the DFT calculation in all the structures.
On the other hand, the MLPs with low RMS errors derive the temperature dependence of the thermal expansion accurately in all the structures.
The accurate prediction of the thermal expansion indicates that the MLPs with low RMS errors can accurately evaluate the volume dependence of the whole range of phonon frequencies.

\begin{figure}[tbp]
\includegraphics[clip,width=\linewidth]{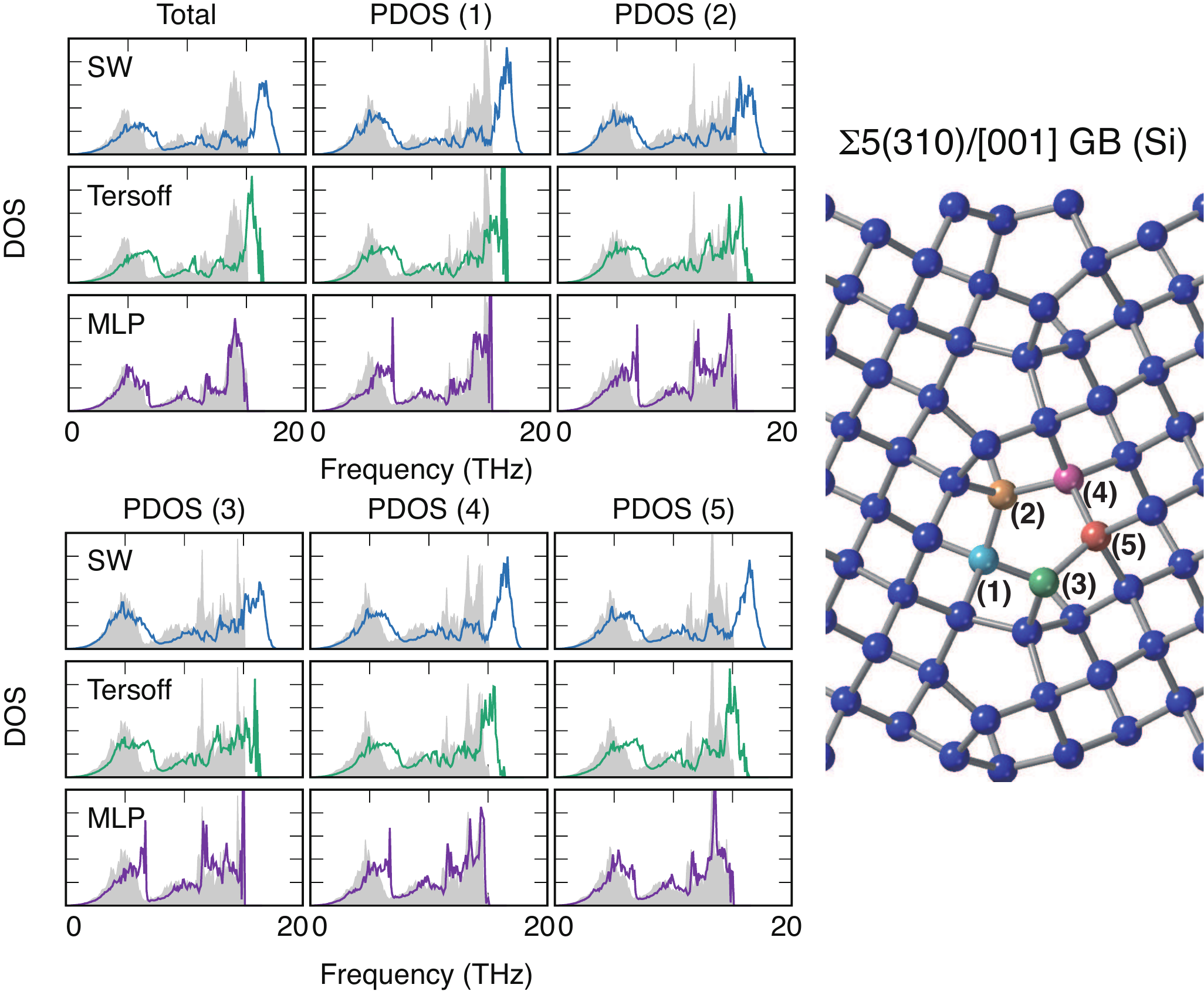}
\caption{
Projected phonon DOS for sites on the grain boundary plane of the $\Sigma$5(310) grain boundary structure predicted using the SW potential\cite{PhysRevB.31.5262}, Tersoff potential\cite{PhysRevB.37.6991}, and polynomial MLP\cite{FUJII2022111137}. 
The total phonon DOS for the grain boundary structures is also shown. 
The shaded region indicates the phonon DOS computed using the DFT calculation.
}
\label{tutorial-2022:Fig-Si-gb-phonon}
\end{figure}

Fujii and Seko compared the projected phonon DOS for sites on the grain boundary plane of $\Sigma$5(310) and $\Sigma$3(112) grain boundary structures \cite{FUJII2022111137} with those computed using the Stillinger--Weber (SW) potential \cite{PhysRevB.31.5262}, Tersoff potential \cite{PhysRevB.37.6991}, and the DFT calculation.
Figure \ref{tutorial-2022:Fig-Si-gb-phonon} shows the projected phonon DOS for five sites on the grain boundary plane of the $\Sigma$5(310) grain boundary.
The SW and Tersoff potentials tend to overestimate the phonon frequencies of the grain boundary structures.
On the other hand, the total and projected phonon DOSs computed using the polynomial MLP are very close to those computed using the DFT calculation. 

In Ref. \onlinecite{FUJII2022111137}, the temperature dependence of the lattice thermal conductivity of the diamond-type structure predicted using the polynomial MLP in silicon was also predicted. 
The lattice thermal conductivity was calculated by using the finite displacement method and solving the Boltzmann transport equation with the single-mode relaxation-time approximation implemented in \textsc{phono3py} \cite{PhysRevB.91.094306}.
To reconstruct the lattice thermal conductivity, the third-order force constants must be accurately predicted using the MLPs.
As found in Fig. \ref{tutorial-2022:Fig-Si-convergence} and the Appendix of Ref. \onlinecite{FUJII2022111137}, the lattice thermal conductivity predicted by the polynomial MLP is consistent with that predicted using the DFT calculation, which is different from the cases of the SW potential and Tersoff potential.

\subsection{Properties on crystallographic defects}

\begin{figure}[tbp]
\includegraphics[clip,width=0.7\linewidth]{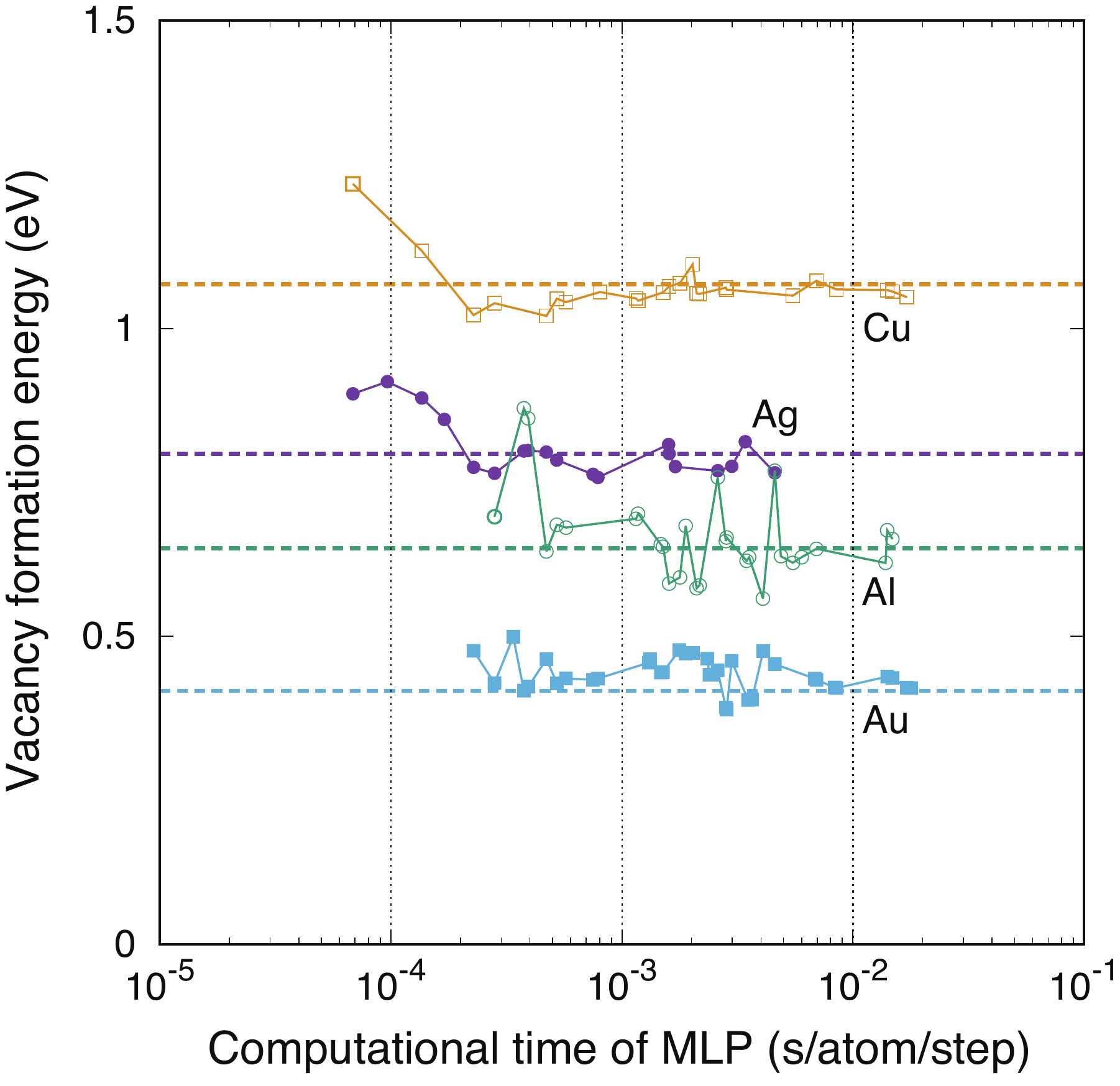}
\caption{
Dependence of the vacancy formation energy for the fcc structure on the selection of the MLP in elemental Ag, Al, Au, and Cu.
The broken lines indicate the vacancy formation energy computed by the DFT calculation.
Here, the vacancy formation energy is estimated using the supercell constructed by the $3\times3\times3$ expansion of the fcc conventional unit cell.
}
\label{tutorial-2022:Fig-fcc-vacancy}
\end{figure}

In the datasets used for developing the current polynomial MLPs, no structures in which crystal defects are explicitly introduced are included, although many of them are generated by introducing large lattice distortions and large atomic displacements into structure generators.
Therefore, the predictive power for properties related to simple crystallographic defects should be examined before performing large-scale simulations with some crystallographic defects.
Firstly, the predictive power for the vacancy formation energy is demonstrated.
Figure \ref{tutorial-2022:Fig-fcc-vacancy} shows the dependence of the vacancy formation energy for the fcc structure on the selection of the MLP in elemental Ag, Al, Au, and Cu.
The vacancy formation energy tends to converge as the model complexity increases, which is similar to the other properties shown earlier.
However, small scattering of the predicted vacancy formation energy compared with the other primitive bulk properties is recognized in Fig. \ref{tutorial-2022:Fig-fcc-vacancy}, especially in elemental Al having the largest RMS error among the four metals as shown in Fig. \ref{tutorial-2022:Fig-pareto-elements-alloy}.
Since such a small deviation from the DFT calculation should be improved by a small modification of the regression coefficients, the small scattering may be reduced by estimating MLPs with a small number of additional data including structures with some vacancies or structures with low-density local regions.
Nonetheless, the converged vacancy formation energies of the polynomial MLP in the four metals are similar to those predicted by the DFT calculation.

\begin{figure}[tbp]
\includegraphics[clip,width=0.8\linewidth]{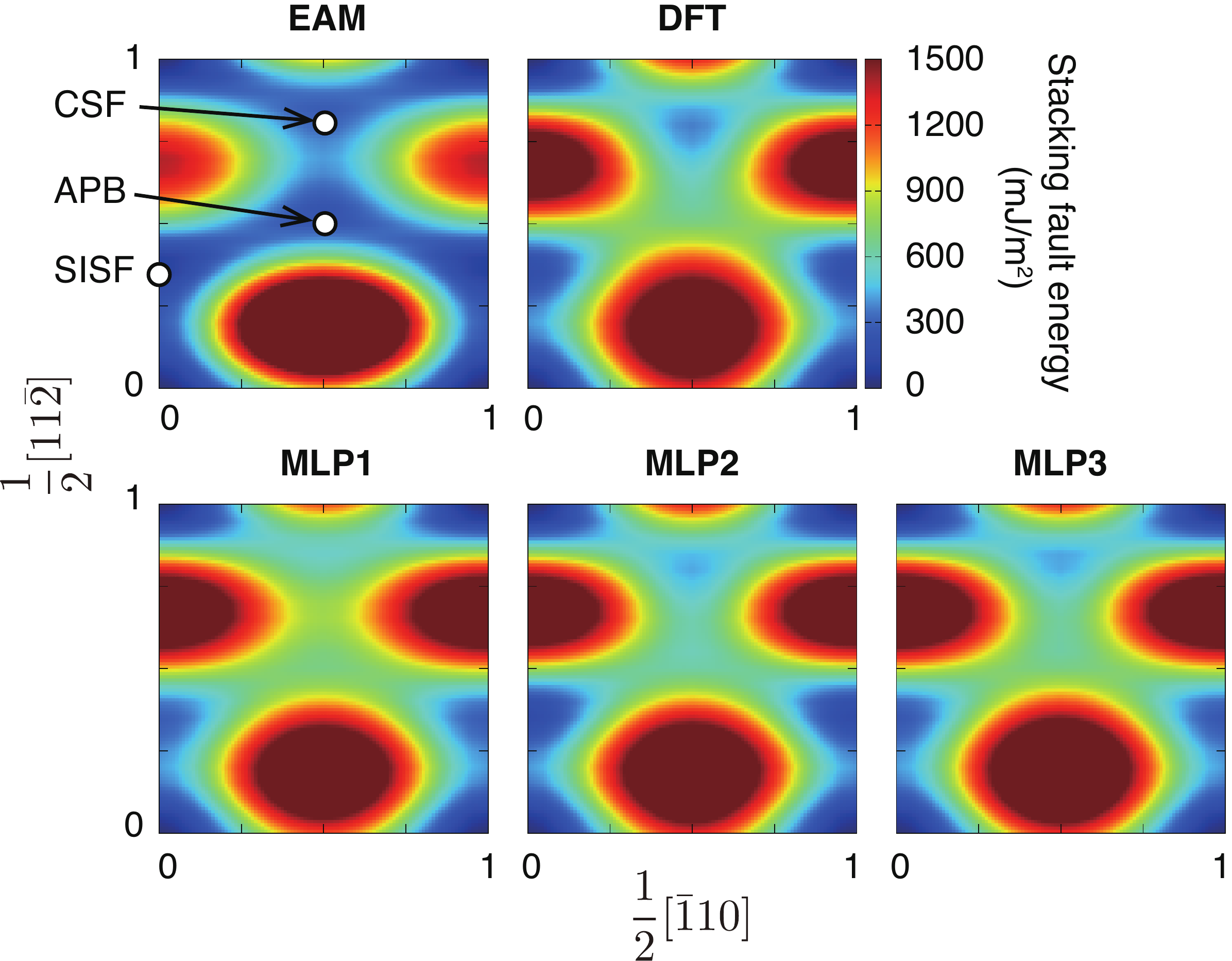}
\caption{
GSFE surfaces in $\gamma$-TiAl obtained using the EAM potential\cite{PhysRevB.68.024102}, three MLPs, and DFT calculation.
MLP1, MLP2, and MLP3 show RMS errors of 4.5, 2.1, and 1.7 meV/atom, respectively.
}
\label{tutorial-2022:Fig-Ti-Al-gsfe-map}
\end{figure}

In Ref. \onlinecite{PhysRevB.102.174104}, the predictive power for the stacking fault energy was examined in $\gamma$-TiAl.
The stacking fault energy can be defined for not only these special stacking faults but also other general stacking faults defined by displacement vectors.
A collection of the excessive energies of the general stacking faults comprises a generalized stacking fault energy (GSFE) surface.
The displacement vector identifying the tilt of the supercell for a general stacking fault is given by
\begin{equation}
\bm{b} = u \left[ \frac{1}{2} \langle \bar110 \rangle \right]
+ v\left[ \frac{1}{2} \langle 11\bar2 \rangle \right],
\end{equation}
where $u$ and $v$ denote the fractional coordinates defined by the vectors $\langle\bar110\rangle/2$ and $\langle11\bar2\rangle/2$, respectively.
Figure \ref{tutorial-2022:Fig-Ti-Al-gsfe-map} shows the GSFE surfaces in $\gamma$-TiAl predicted using the EAM potential, three MLPs, and DFT calculation.
The GSFE surfaces predicted using the MLPs and DFT calculation are similar, whereas that predicted using the EAM potential is different from that predicted by the DFT calculation.
Thus, the current polynomial MLPs should have high predictive power for the stacking faults and related properties.

\begin{figure}[tbp]
\includegraphics[clip,width=\linewidth]{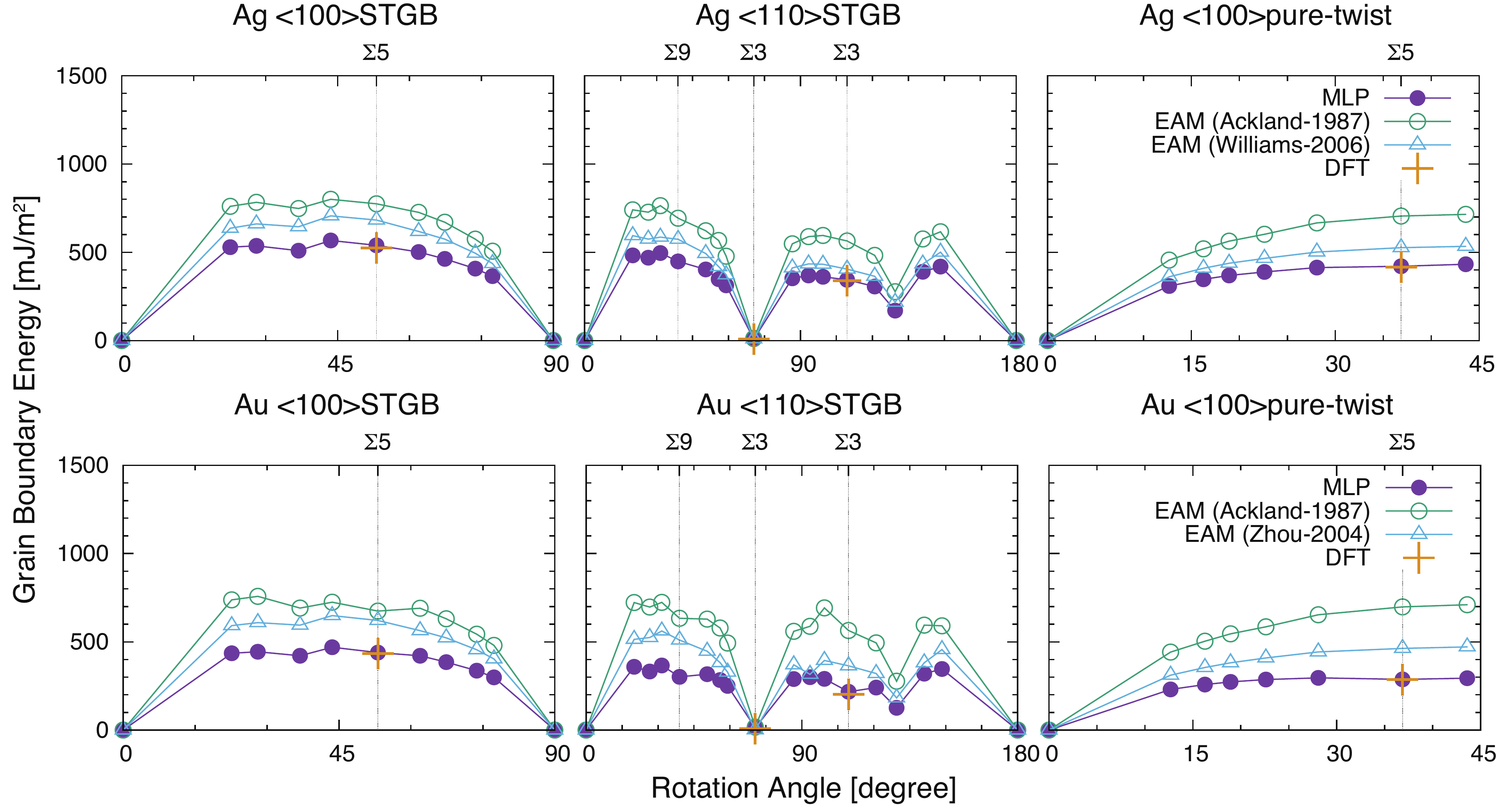}
\caption{
Rotation angle dependence of the grain boundary energies for $\langle 100 \rangle$ STGBs, $\langle 110 \rangle$ STGBs, and $\langle 100 \rangle$ pure-twist grain boundaries in fcc elemental Ag and Au predicted using the Pareto-optimal polynomial MLPs \cite{PhysRevMaterials.4.123607}.
The grain boundary energies predicted using EAM potentials \cite{doi:10.1080/01418618708204485,Williams_2006,PhysRevB.69.144113} are also shown.
}
\label{tutorial-2022:Fig-fcc-gb-energy}
\end{figure}

Nishiyama et al. examined the predictive power of polynomial MLPs for grain boundary properties by systematically evaluating the grain boundary energy for $\langle 100 \rangle$ symmetric tilt grain boundaries (STGBs), $\langle 110 \rangle$ STGBs, and $\langle 100 \rangle$ pure-twist grain boundaries in the fcc elemental metals Ag, Al, Au, Cu, Pd, and Pt \cite{PhysRevMaterials.4.123607}. 
In every elemental metal, the values of the grain boundary energy computed using the MLP are consistent with those computed by DFT calculation as shown in Fig. \ref{tutorial-2022:Fig-fcc-gb-energy}. 
It is worth emphasizing that the training datasets used to develop the MLPs contain no grain boundary structures. 
Moreover, Fujii and Seko compared the grain boundary energies for several grain boundaries predicted using polynomial MLPs with those obtained by the DFT calculation in elemental Si\cite{FUJII2022111137}. 
Their convergence behaviors with respect to the computational time of the MLP are shown in Fig. \ref{tutorial-2022:Fig-Si-convergence}.
Similarly to in the fcc elemental metals, the converged grain boundary energies are consistent with the DFT grain boundary energy.

\section{Conclusion}
\label{tutorial-2022:Sec-conclusion}
This paper has overviewed the formulation of the polynomial MLP, the procedures for estimating the polynomial MLP, and the systematic development of the polynomial MLPs for various elemental and binary alloy systems.
The polynomial MLPs are formulated using polynomial invariants for the O(3) group enumerated from order parameters in terms of products of radial and spherical harmonic functions.
Combined with a dataset constructed by systematic DFT calculations for a wide variety of structures, the polynomial MLPs have high predictive power for many structures and properties.

Currently, the polynomial MLPs with different trade-offs between accuracy and computational efficiency for 164 elemental and binary alloy systems are available in the repository from the website \cite{MachineLearningPotentialRepository}.
The number of MLP entries in the repository is continuously increasing.
In particular, MLPs with high computational cost performance in the repository, showing high computational efficiency without increasing the prediction error, should help users to perform accurate and fast atomistic simulations for most practical purposes.

\begin{acknowledgments}
This work was supported by a Grant-in-Aid for Scientific Research (B) (Grant Number 22H01756) and a Grant-in-Aid for Scientific Research on Innovative Areas (Grant Number 19H05787) from the Japan Society for the Promotion of Science (JSPS).
\end{acknowledgments}

\bibliography{tutorial}

\begin{thebibliography}{64}%
\makeatletter
\providecommand \@ifxundefined [1]{%
 \@ifx{#1\undefined}
}%
\providecommand \@ifnum [1]{%
 \ifnum #1\expandafter \@firstoftwo
 \else \expandafter \@secondoftwo
 \fi
}%
\providecommand \@ifx [1]{%
 \ifx #1\expandafter \@firstoftwo
 \else \expandafter \@secondoftwo
 \fi
}%
\providecommand \natexlab [1]{#1}%
\providecommand \enquote  [1]{``#1''}%
\providecommand \bibnamefont  [1]{#1}%
\providecommand \bibfnamefont [1]{#1}%
\providecommand \citenamefont [1]{#1}%
\providecommand \href@noop [0]{\@secondoftwo}%
\providecommand \href [0]{\begingroup \@sanitize@url \@href}%
\providecommand \@href[1]{\@@startlink{#1}\@@href}%
\providecommand \@@href[1]{\endgroup#1\@@endlink}%
\providecommand \@sanitize@url [0]{\catcode `\\12\catcode `\$12\catcode
  `\&12\catcode `\#12\catcode `\^12\catcode `\_12\catcode `\%12\relax}%
\providecommand \@@startlink[1]{}%
\providecommand \@@endlink[0]{}%
\providecommand \url  [0]{\begingroup\@sanitize@url \@url }%
\providecommand \@url [1]{\endgroup\@href {#1}{\urlprefix }}%
\providecommand \urlprefix  [0]{URL }%
\providecommand \Eprint [0]{\href }%
\providecommand \doibase [0]{http://dx.doi.org/}%
\providecommand \selectlanguage [0]{\@gobble}%
\providecommand \bibinfo  [0]{\@secondoftwo}%
\providecommand \bibfield  [0]{\@secondoftwo}%
\providecommand \translation [1]{[#1]}%
\providecommand \BibitemOpen [0]{}%
\providecommand \bibitemStop [0]{}%
\providecommand \bibitemNoStop [0]{.\EOS\space}%
\providecommand \EOS [0]{\spacefactor3000\relax}%
\providecommand \BibitemShut  [1]{\csname bibitem#1\endcsname}%
\let\auto@bib@innerbib\@empty
\bibitem [{Mac()}]{MachineLearningPotentialRepository}%
  \BibitemOpen
  \href {https://sekocha.github.io} {}\bibinfo {note} {{A. Seko}, {Polynomial}
  {Machine} {Learning} {Potential} {Repository} at {Kyoto} {University},
  \url{https://sekocha.github.io}}\BibitemShut {NoStop}%
\bibitem [{\citenamefont {Lorenz}, \citenamefont {Gro{\ss}},\ and\
  \citenamefont {Scheffler}(2004)}]{Lorenz2004210}%
  \BibitemOpen
  \bibfield  {author} {\bibinfo {author} {\bibfnamefont {S.}~\bibnamefont
  {Lorenz}}, \bibinfo {author} {\bibfnamefont {A.}~\bibnamefont {Gro{\ss}}}, \
  and\ \bibinfo {author} {\bibfnamefont {M.}~\bibnamefont {Scheffler}},\ }\href
  {\doibase http://dx.doi.org/10.1016/j.cplett.2004.07.076} {\bibfield
  {journal} {\bibinfo  {journal} {Chem. Phys. Lett.}\ }\textbf {\bibinfo
  {volume} {395}},\ \bibinfo {pages} {210 } (\bibinfo {year}
  {2004})}\BibitemShut {NoStop}%
\bibitem [{\citenamefont {Behler}\ and\ \citenamefont
  {Parrinello}(2007)}]{behler2007generalized}%
  \BibitemOpen
  \bibfield  {author} {\bibinfo {author} {\bibfnamefont {J.}~\bibnamefont
  {Behler}}\ and\ \bibinfo {author} {\bibfnamefont {M.}~\bibnamefont
  {Parrinello}},\ }\href@noop {} {\bibfield  {journal} {\bibinfo  {journal}
  {Phys. Rev. Lett.}\ }\textbf {\bibinfo {volume} {98}},\ \bibinfo {pages}
  {146401} (\bibinfo {year} {2007})}\BibitemShut {NoStop}%
\bibitem [{\citenamefont {Bart{\'o}k}\ \emph {et~al.}(2010)\citenamefont
  {Bart{\'o}k}, \citenamefont {Payne}, \citenamefont {Kondor},\ and\
  \citenamefont {Cs{\'a}nyi}}]{bartok2010gaussian}%
  \BibitemOpen
  \bibfield  {author} {\bibinfo {author} {\bibfnamefont {A.~P.}\ \bibnamefont
  {Bart{\'o}k}}, \bibinfo {author} {\bibfnamefont {M.~C.}\ \bibnamefont
  {Payne}}, \bibinfo {author} {\bibfnamefont {R.}~\bibnamefont {Kondor}}, \
  and\ \bibinfo {author} {\bibfnamefont {G.}~\bibnamefont {Cs{\'a}nyi}},\
  }\href@noop {} {\bibfield  {journal} {\bibinfo  {journal} {Phys. Rev. Lett.}\
  }\textbf {\bibinfo {volume} {104}},\ \bibinfo {pages} {136403} (\bibinfo
  {year} {2010})}\BibitemShut {NoStop}%
\bibitem [{\citenamefont {Behler}(2011)}]{behler2011atom}%
  \BibitemOpen
  \bibfield  {author} {\bibinfo {author} {\bibfnamefont {J.}~\bibnamefont
  {Behler}},\ }\href@noop {} {\bibfield  {journal} {\bibinfo  {journal} {J.
  Chem. Phys.}\ }\textbf {\bibinfo {volume} {134}},\ \bibinfo {pages} {074106}
  (\bibinfo {year} {2011})}\BibitemShut {NoStop}%
\bibitem [{\citenamefont {Han}\ \emph {et~al.}(2018)\citenamefont {Han},
  \citenamefont {Zhang}, \citenamefont {Car},\ and\ \citenamefont
  {Weinan}}]{han2017deep}%
  \BibitemOpen
  \bibfield  {author} {\bibinfo {author} {\bibfnamefont {J.}~\bibnamefont
  {Han}}, \bibinfo {author} {\bibfnamefont {L.}~\bibnamefont {Zhang}}, \bibinfo
  {author} {\bibfnamefont {R.}~\bibnamefont {Car}}, \ and\ \bibinfo {author}
  {\bibfnamefont {E.}~\bibnamefont {Weinan}},\ }\href@noop {} {\bibfield
  {journal} {\bibinfo  {journal} {Commun. Comput. Phys.}\ }\textbf {\bibinfo
  {volume} {23}},\ \bibinfo {pages} {629 } (\bibinfo {year}
  {2018})}\BibitemShut {NoStop}%
\bibitem [{\citenamefont {Artrith}\ and\ \citenamefont
  {Urban}(2016)}]{258c531ae5de4f5699e2eec2de51c84f}%
  \BibitemOpen
  \bibfield  {author} {\bibinfo {author} {\bibfnamefont {N.}~\bibnamefont
  {Artrith}}\ and\ \bibinfo {author} {\bibfnamefont {A.}~\bibnamefont
  {Urban}},\ }\href {\doibase 10.1016/j.commatsci.2015.11.047} {\bibfield
  {journal} {\bibinfo  {journal} {Comput. Mater. Sci.}\ }\textbf {\bibinfo
  {volume} {114}},\ \bibinfo {pages} {135} (\bibinfo {year}
  {2016})}\BibitemShut {NoStop}%
\bibitem [{\citenamefont {Artrith}, \citenamefont {Urban},\ and\ \citenamefont
  {Ceder}(2017)}]{PhysRevB.96.014112}%
  \BibitemOpen
  \bibfield  {author} {\bibinfo {author} {\bibfnamefont {N.}~\bibnamefont
  {Artrith}}, \bibinfo {author} {\bibfnamefont {A.}~\bibnamefont {Urban}}, \
  and\ \bibinfo {author} {\bibfnamefont {G.}~\bibnamefont {Ceder}},\ }\href
  {\doibase 10.1103/PhysRevB.96.014112} {\bibfield  {journal} {\bibinfo
  {journal} {Phys. Rev. B}\ }\textbf {\bibinfo {volume} {96}},\ \bibinfo
  {pages} {014112} (\bibinfo {year} {2017})}\BibitemShut {NoStop}%
\bibitem [{\citenamefont {Szlachta}, \citenamefont {Bart\'ok},\ and\
  \citenamefont {Cs\'anyi}(2014)}]{PhysRevB.90.104108}%
  \BibitemOpen
  \bibfield  {author} {\bibinfo {author} {\bibfnamefont {W.~J.}\ \bibnamefont
  {Szlachta}}, \bibinfo {author} {\bibfnamefont {A.~P.}\ \bibnamefont
  {Bart\'ok}}, \ and\ \bibinfo {author} {\bibfnamefont {G.}~\bibnamefont
  {Cs\'anyi}},\ }\href {\doibase 10.1103/PhysRevB.90.104108} {\bibfield
  {journal} {\bibinfo  {journal} {Phys. Rev. B}\ }\textbf {\bibinfo {volume}
  {90}},\ \bibinfo {pages} {104108} (\bibinfo {year} {2014})}\BibitemShut
  {NoStop}%
\bibitem [{\citenamefont {Bart\'ok}\ \emph {et~al.}(2018)\citenamefont
  {Bart\'ok}, \citenamefont {Kermode}, \citenamefont {Bernstein},\ and\
  \citenamefont {Cs\'anyi}}]{PhysRevX.8.041048}%
  \BibitemOpen
  \bibfield  {author} {\bibinfo {author} {\bibfnamefont {A.~P.}\ \bibnamefont
  {Bart\'ok}}, \bibinfo {author} {\bibfnamefont {J.}~\bibnamefont {Kermode}},
  \bibinfo {author} {\bibfnamefont {N.}~\bibnamefont {Bernstein}}, \ and\
  \bibinfo {author} {\bibfnamefont {G.}~\bibnamefont {Cs\'anyi}},\ }\href
  {\doibase 10.1103/PhysRevX.8.041048} {\bibfield  {journal} {\bibinfo
  {journal} {Phys. Rev. X}\ }\textbf {\bibinfo {volume} {8}},\ \bibinfo {pages}
  {041048} (\bibinfo {year} {2018})}\BibitemShut {NoStop}%
\bibitem [{\citenamefont {Li}, \citenamefont {Kermode},\ and\ \citenamefont
  {De~Vita}(2015)}]{PhysRevLett.114.096405}%
  \BibitemOpen
  \bibfield  {author} {\bibinfo {author} {\bibfnamefont {Z.}~\bibnamefont
  {Li}}, \bibinfo {author} {\bibfnamefont {J.~R.}\ \bibnamefont {Kermode}}, \
  and\ \bibinfo {author} {\bibfnamefont {A.}~\bibnamefont {De~Vita}},\ }\href
  {\doibase 10.1103/PhysRevLett.114.096405} {\bibfield  {journal} {\bibinfo
  {journal} {Phys. Rev. Lett.}\ }\textbf {\bibinfo {volume} {114}},\ \bibinfo
  {pages} {096405} (\bibinfo {year} {2015})}\BibitemShut {NoStop}%
\bibitem [{\citenamefont {Glielmo}, \citenamefont {Sollich},\ and\
  \citenamefont {De~Vita}(2017)}]{PhysRevB.95.214302}%
  \BibitemOpen
  \bibfield  {author} {\bibinfo {author} {\bibfnamefont {A.}~\bibnamefont
  {Glielmo}}, \bibinfo {author} {\bibfnamefont {P.}~\bibnamefont {Sollich}}, \
  and\ \bibinfo {author} {\bibfnamefont {A.}~\bibnamefont {De~Vita}},\ }\href
  {\doibase 10.1103/PhysRevB.95.214302} {\bibfield  {journal} {\bibinfo
  {journal} {Phys. Rev. B}\ }\textbf {\bibinfo {volume} {95}},\ \bibinfo
  {pages} {214302} (\bibinfo {year} {2017})}\BibitemShut {NoStop}%
\bibitem [{\citenamefont {Seko}, \citenamefont {Takahashi},\ and\ \citenamefont
  {Tanaka}(2014)}]{PhysRevB.90.024101}%
  \BibitemOpen
  \bibfield  {author} {\bibinfo {author} {\bibfnamefont {A.}~\bibnamefont
  {Seko}}, \bibinfo {author} {\bibfnamefont {A.}~\bibnamefont {Takahashi}}, \
  and\ \bibinfo {author} {\bibfnamefont {I.}~\bibnamefont {Tanaka}},\ }\href
  {\doibase 10.1103/PhysRevB.90.024101} {\bibfield  {journal} {\bibinfo
  {journal} {Phys. Rev. B}\ }\textbf {\bibinfo {volume} {90}},\ \bibinfo
  {pages} {024101} (\bibinfo {year} {2014})}\BibitemShut {NoStop}%
\bibitem [{\citenamefont {Seko}, \citenamefont {Takahashi},\ and\ \citenamefont
  {Tanaka}(2015)}]{PhysRevB.92.054113}%
  \BibitemOpen
  \bibfield  {author} {\bibinfo {author} {\bibfnamefont {A.}~\bibnamefont
  {Seko}}, \bibinfo {author} {\bibfnamefont {A.}~\bibnamefont {Takahashi}}, \
  and\ \bibinfo {author} {\bibfnamefont {I.}~\bibnamefont {Tanaka}},\ }\href
  {\doibase 10.1103/PhysRevB.92.054113} {\bibfield  {journal} {\bibinfo
  {journal} {Phys. Rev. B}\ }\textbf {\bibinfo {volume} {92}},\ \bibinfo
  {pages} {054113} (\bibinfo {year} {2015})}\BibitemShut {NoStop}%
\bibitem [{\citenamefont {Takahashi}, \citenamefont {Seko},\ and\ \citenamefont
  {Tanaka}(2017)}]{PhysRevMaterials.1.063801}%
  \BibitemOpen
  \bibfield  {author} {\bibinfo {author} {\bibfnamefont {A.}~\bibnamefont
  {Takahashi}}, \bibinfo {author} {\bibfnamefont {A.}~\bibnamefont {Seko}}, \
  and\ \bibinfo {author} {\bibfnamefont {I.}~\bibnamefont {Tanaka}},\ }\href
  {\doibase 10.1103/PhysRevMaterials.1.063801} {\bibfield  {journal} {\bibinfo
  {journal} {Phys. Rev. Mater.}\ }\textbf {\bibinfo {volume} {1}},\ \bibinfo
  {pages} {063801} (\bibinfo {year} {2017})}\BibitemShut {NoStop}%
\bibitem [{\citenamefont {Thompson}\ \emph {et~al.}(2015)\citenamefont
  {Thompson}, \citenamefont {Swiler}, \citenamefont {Trott}, \citenamefont
  {Foiles},\ and\ \citenamefont {Tucker}}]{Thompson2015316}%
  \BibitemOpen
  \bibfield  {author} {\bibinfo {author} {\bibfnamefont {A.}~\bibnamefont
  {Thompson}}, \bibinfo {author} {\bibfnamefont {L.}~\bibnamefont {Swiler}},
  \bibinfo {author} {\bibfnamefont {C.}~\bibnamefont {Trott}}, \bibinfo
  {author} {\bibfnamefont {S.}~\bibnamefont {Foiles}}, \ and\ \bibinfo {author}
  {\bibfnamefont {G.}~\bibnamefont {Tucker}},\ }\href {\doibase
  https://doi.org/10.1016/j.jcp.2014.12.018} {\bibfield  {journal} {\bibinfo
  {journal} {J. Comput. Phys.}\ }\textbf {\bibinfo {volume} {285}},\ \bibinfo
  {pages} {316 } (\bibinfo {year} {2015})}\BibitemShut {NoStop}%
\bibitem [{\citenamefont {Wood}\ and\ \citenamefont
  {Thompson}(2018{\natexlab{a}})}]{wood2018extending}%
  \BibitemOpen
  \bibfield  {author} {\bibinfo {author} {\bibfnamefont {M.~A.}\ \bibnamefont
  {Wood}}\ and\ \bibinfo {author} {\bibfnamefont {A.~P.}\ \bibnamefont
  {Thompson}},\ }\href@noop {} {\bibfield  {journal} {\bibinfo  {journal} {J.
  Chem. Phys.}\ }\textbf {\bibinfo {volume} {148}},\ \bibinfo {pages} {241721}
  (\bibinfo {year} {2018}{\natexlab{a}})}\BibitemShut {NoStop}%
\bibitem [{\citenamefont {Chen}\ \emph {et~al.}(2017)\citenamefont {Chen},
  \citenamefont {Deng}, \citenamefont {Tran}, \citenamefont {Tang},
  \citenamefont {Chu},\ and\ \citenamefont {Ong}}]{PhysRevMaterials.1.043603}%
  \BibitemOpen
  \bibfield  {author} {\bibinfo {author} {\bibfnamefont {C.}~\bibnamefont
  {Chen}}, \bibinfo {author} {\bibfnamefont {Z.}~\bibnamefont {Deng}}, \bibinfo
  {author} {\bibfnamefont {R.}~\bibnamefont {Tran}}, \bibinfo {author}
  {\bibfnamefont {H.}~\bibnamefont {Tang}}, \bibinfo {author} {\bibfnamefont
  {I.-H.}\ \bibnamefont {Chu}}, \ and\ \bibinfo {author} {\bibfnamefont
  {S.~P.}\ \bibnamefont {Ong}},\ }\href {\doibase
  10.1103/PhysRevMaterials.1.043603} {\bibfield  {journal} {\bibinfo  {journal}
  {Phys. Rev. Mater.}\ }\textbf {\bibinfo {volume} {1}},\ \bibinfo {pages}
  {043603} (\bibinfo {year} {2017})}\BibitemShut {NoStop}%
\bibitem [{\citenamefont {Shapeev}(2016)}]{doi-10.1137-15M1054183}%
  \BibitemOpen
  \bibfield  {author} {\bibinfo {author} {\bibfnamefont {A.~V.}\ \bibnamefont
  {Shapeev}},\ }\href {\doibase 10.1137/15M1054183} {\bibfield  {journal}
  {\bibinfo  {journal} {Multiscale Model. Simul.}\ }\textbf {\bibinfo {volume}
  {14}},\ \bibinfo {pages} {1153} (\bibinfo {year} {2016})}\BibitemShut
  {NoStop}%
\bibitem [{\citenamefont {Deringer}, \citenamefont {Pickard},\ and\
  \citenamefont {Cs\'anyi}(2018)}]{PhysRevLett.120.156001}%
  \BibitemOpen
  \bibfield  {author} {\bibinfo {author} {\bibfnamefont {V.~L.}\ \bibnamefont
  {Deringer}}, \bibinfo {author} {\bibfnamefont {C.~J.}\ \bibnamefont
  {Pickard}}, \ and\ \bibinfo {author} {\bibfnamefont {G.}~\bibnamefont
  {Cs\'anyi}},\ }\href {\doibase 10.1103/PhysRevLett.120.156001} {\bibfield
  {journal} {\bibinfo  {journal} {Phys. Rev. Lett.}\ }\textbf {\bibinfo
  {volume} {120}},\ \bibinfo {pages} {156001} (\bibinfo {year}
  {2018})}\BibitemShut {NoStop}%
\bibitem [{\citenamefont {Podryabinkin}\ \emph {et~al.}(2018)\citenamefont
  {Podryabinkin}, \citenamefont {Tikhonov}, \citenamefont {Shapeev},\ and\
  \citenamefont {Oganov}}]{podryabinkin2018accelerating}%
  \BibitemOpen
  \bibfield  {author} {\bibinfo {author} {\bibfnamefont {E.~V.}\ \bibnamefont
  {Podryabinkin}}, \bibinfo {author} {\bibfnamefont {E.~V.}\ \bibnamefont
  {Tikhonov}}, \bibinfo {author} {\bibfnamefont {A.~V.}\ \bibnamefont
  {Shapeev}}, \ and\ \bibinfo {author} {\bibfnamefont {A.~R.}\ \bibnamefont
  {Oganov}},\ }\href@noop {} {\bibfield  {journal} {\bibinfo  {journal} {arXiv
  preprint arXiv:1802.07605}\ } (\bibinfo {year} {2018})}\BibitemShut {NoStop}%
\bibitem [{\citenamefont {Gubaev}\ \emph {et~al.}(2019)\citenamefont {Gubaev},
  \citenamefont {Podryabinkin}, \citenamefont {Hart},\ and\ \citenamefont
  {Shapeev}}]{GUBAEV2019148}%
  \BibitemOpen
  \bibfield  {author} {\bibinfo {author} {\bibfnamefont {K.}~\bibnamefont
  {Gubaev}}, \bibinfo {author} {\bibfnamefont {E.~V.}\ \bibnamefont
  {Podryabinkin}}, \bibinfo {author} {\bibfnamefont {G.~L.}\ \bibnamefont
  {Hart}}, \ and\ \bibinfo {author} {\bibfnamefont {A.~V.}\ \bibnamefont
  {Shapeev}},\ }\href {\doibase
  https://doi.org/10.1016/j.commatsci.2018.09.031} {\bibfield  {journal}
  {\bibinfo  {journal} {Comput. Mater. Sci.}\ }\textbf {\bibinfo {volume}
  {156}},\ \bibinfo {pages} {148 } (\bibinfo {year} {2019})}\BibitemShut
  {NoStop}%
\bibitem [{\citenamefont {Mueller}, \citenamefont {Hernandez},\ and\
  \citenamefont {Wang}(2020)}]{doi:10.1063/1.5126336}%
  \BibitemOpen
  \bibfield  {author} {\bibinfo {author} {\bibfnamefont {T.}~\bibnamefont
  {Mueller}}, \bibinfo {author} {\bibfnamefont {A.}~\bibnamefont {Hernandez}},
  \ and\ \bibinfo {author} {\bibfnamefont {C.}~\bibnamefont {Wang}},\ }\href
  {\doibase 10.1063/1.5126336} {\bibfield  {journal} {\bibinfo  {journal} {J.
  Chem. Phys.}\ }\textbf {\bibinfo {volume} {152}},\ \bibinfo {pages} {050902}
  (\bibinfo {year} {2020})}\BibitemShut {NoStop}%
\bibitem [{\citenamefont {Freitas}\ and\ \citenamefont
  {Cao}(2022)}]{Freitas2022}%
  \BibitemOpen
  \bibfield  {author} {\bibinfo {author} {\bibfnamefont {R.}~\bibnamefont
  {Freitas}}\ and\ \bibinfo {author} {\bibfnamefont {Y.}~\bibnamefont {Cao}},\
  }\href {\doibase 10.1557/s43579-022-00221-5} {\bibfield  {journal} {\bibinfo
  {journal} {MRS Commun.}\ } (\bibinfo {year} {2022}),\
  10.1557/s43579-022-00221-5}\BibitemShut {NoStop}%
\bibitem [{NIST Interatomic Potentials
  Repository()}]{interatomicPotentialRepository}%
  \BibitemOpen
  NIST Interatomic Potentials Repository,\ \href@noop {} {}\bibinfo
  {howpublished} {\url{http://www.ctcms.nist.gov/potentials}}\BibitemShut
  {NoStop}%
\bibitem [{E. Tadmor, R. Elliott, J. Sethna, R. Miller, and C. Becker,
  Knowledgebase of interatomic models (KIM)(2011)}]{KIMproject}%
  \BibitemOpen
  E. Tadmor, R. Elliott, J. Sethna, R. Miller, and C. Becker, Knowledgebase of
  interatomic models (KIM),\ \href@noop {} {}\bibinfo {howpublished}
  {\url{https://openkim.org}} (\bibinfo {year} {2011})\BibitemShut {NoStop}%
\bibitem [{\citenamefont {Seko}, \citenamefont {Togo},\ and\ \citenamefont
  {Tanaka}(2019)}]{PhysRevB.99.214108}%
  \BibitemOpen
  \bibfield  {author} {\bibinfo {author} {\bibfnamefont {A.}~\bibnamefont
  {Seko}}, \bibinfo {author} {\bibfnamefont {A.}~\bibnamefont {Togo}}, \ and\
  \bibinfo {author} {\bibfnamefont {I.}~\bibnamefont {Tanaka}},\ }\href
  {\doibase 10.1103/PhysRevB.99.214108} {\bibfield  {journal} {\bibinfo
  {journal} {Phys. Rev. B}\ }\textbf {\bibinfo {volume} {99}},\ \bibinfo
  {pages} {214108} (\bibinfo {year} {2019})}\BibitemShut {NoStop}%
\bibitem [{\citenamefont {Hernandez}\ \emph {et~al.}(2019)\citenamefont
  {Hernandez}, \citenamefont {Balasubramanian}, \citenamefont {Yuan},
  \citenamefont {Mason},\ and\ \citenamefont {Mueller}}]{hernandez2019fast}%
  \BibitemOpen
  \bibfield  {author} {\bibinfo {author} {\bibfnamefont {A.}~\bibnamefont
  {Hernandez}}, \bibinfo {author} {\bibfnamefont {A.}~\bibnamefont
  {Balasubramanian}}, \bibinfo {author} {\bibfnamefont {F.}~\bibnamefont
  {Yuan}}, \bibinfo {author} {\bibfnamefont {S.~A.}\ \bibnamefont {Mason}}, \
  and\ \bibinfo {author} {\bibfnamefont {T.}~\bibnamefont {Mueller}},\
  }\href@noop {} {\bibfield  {journal} {\bibinfo  {journal} {npj Comput.
  Mater.}\ }\textbf {\bibinfo {volume} {5}},\ \bibinfo {pages} {1} (\bibinfo
  {year} {2019})}\BibitemShut {NoStop}%
\bibitem [{\citenamefont {Zuo}\ \emph {et~al.}(2020)\citenamefont {Zuo},
  \citenamefont {Chen}, \citenamefont {Li}, \citenamefont {Deng}, \citenamefont
  {Chen}, \citenamefont {Behler}, \citenamefont {Cs\'anyi}, \citenamefont
  {Shapeev}, \citenamefont {Thompson}, \citenamefont {Wood},\ and\
  \citenamefont {Ong}}]{doi:10.1021/acs.jpca.9b08723}%
  \BibitemOpen
  \bibfield  {author} {\bibinfo {author} {\bibfnamefont {Y.}~\bibnamefont
  {Zuo}}, \bibinfo {author} {\bibfnamefont {C.}~\bibnamefont {Chen}}, \bibinfo
  {author} {\bibfnamefont {X.}~\bibnamefont {Li}}, \bibinfo {author}
  {\bibfnamefont {Z.}~\bibnamefont {Deng}}, \bibinfo {author} {\bibfnamefont
  {Y.}~\bibnamefont {Chen}}, \bibinfo {author} {\bibfnamefont {J.}~\bibnamefont
  {Behler}}, \bibinfo {author} {\bibfnamefont {G.}~\bibnamefont {Cs\'anyi}},
  \bibinfo {author} {\bibfnamefont {A.~V.}\ \bibnamefont {Shapeev}}, \bibinfo
  {author} {\bibfnamefont {A.~P.}\ \bibnamefont {Thompson}}, \bibinfo {author}
  {\bibfnamefont {M.~A.}\ \bibnamefont {Wood}}, \ and\ \bibinfo {author}
  {\bibfnamefont {S.~P.}\ \bibnamefont {Ong}},\ }\href {\doibase
  10.1021/acs.jpca.9b08723} {\bibfield  {journal} {\bibinfo  {journal} {J.
  Phys. Chem. A}\ }\textbf {\bibinfo {volume} {124}},\ \bibinfo {pages} {731}
  (\bibinfo {year} {2020})},\ \bibinfo {note} {pMID: 31916773}\BibitemShut
  {NoStop}%
\bibitem [{\citenamefont {Thompson}\ \emph {et~al.}(2022)\citenamefont
  {Thompson}, \citenamefont {Aktulga}, \citenamefont {Berger}, \citenamefont
  {Bolintineanu}, \citenamefont {Brown}, \citenamefont {Crozier}, \citenamefont
  {in~'t Veld}, \citenamefont {Kohlmeyer}, \citenamefont {Moore}, \citenamefont
  {Nguyen}, \citenamefont {Shan}, \citenamefont {Stevens}, \citenamefont
  {Tranchida}, \citenamefont {Trott},\ and\ \citenamefont {Plimpton}}]{lammps}%
  \BibitemOpen
  \bibfield  {author} {\bibinfo {author} {\bibfnamefont {A.~P.}\ \bibnamefont
  {Thompson}}, \bibinfo {author} {\bibfnamefont {H.~M.}\ \bibnamefont
  {Aktulga}}, \bibinfo {author} {\bibfnamefont {R.}~\bibnamefont {Berger}},
  \bibinfo {author} {\bibfnamefont {D.~S.}\ \bibnamefont {Bolintineanu}},
  \bibinfo {author} {\bibfnamefont {W.~M.}\ \bibnamefont {Brown}}, \bibinfo
  {author} {\bibfnamefont {P.~S.}\ \bibnamefont {Crozier}}, \bibinfo {author}
  {\bibfnamefont {P.~J.}\ \bibnamefont {in~'t Veld}}, \bibinfo {author}
  {\bibfnamefont {A.}~\bibnamefont {Kohlmeyer}}, \bibinfo {author}
  {\bibfnamefont {S.~G.}\ \bibnamefont {Moore}}, \bibinfo {author}
  {\bibfnamefont {T.~D.}\ \bibnamefont {Nguyen}}, \bibinfo {author}
  {\bibfnamefont {R.}~\bibnamefont {Shan}}, \bibinfo {author} {\bibfnamefont
  {M.~J.}\ \bibnamefont {Stevens}}, \bibinfo {author} {\bibfnamefont
  {J.}~\bibnamefont {Tranchida}}, \bibinfo {author} {\bibfnamefont
  {C.}~\bibnamefont {Trott}}, \ and\ \bibinfo {author} {\bibfnamefont {S.~J.}\
  \bibnamefont {Plimpton}},\ }\href {\doibase 10.1016/j.cpc.2021.108171}
  {\bibfield  {journal} {\bibinfo  {journal} {Comput. Phys. Commun.}\ }\textbf
  {\bibinfo {volume} {271}},\ \bibinfo {pages} {108171} (\bibinfo {year}
  {2022})}\BibitemShut {NoStop}%
\bibitem [{Lam()}]{LammpsPolyMLP}%
  \BibitemOpen
  \href {https://github.com/sekocha/lammps-polymlp-package} {}\bibinfo {note}
  {{A. Seko}, {lammps-polymlp-package},
  \url{https://github.com/sekocha/lammps-polymlp-package}}\BibitemShut
  {NoStop}%
\bibitem [{\citenamefont {Seko}(2020)}]{PhysRevB.102.174104}%
  \BibitemOpen
  \bibfield  {author} {\bibinfo {author} {\bibfnamefont {A.}~\bibnamefont
  {Seko}},\ }\href {\doibase 10.1103/PhysRevB.102.174104} {\bibfield  {journal}
  {\bibinfo  {journal} {Phys. Rev. B}\ }\textbf {\bibinfo {volume} {102}},\
  \bibinfo {pages} {174104} (\bibinfo {year} {2020})}\BibitemShut {NoStop}%
\bibitem [{\citenamefont {Bart{\'o}k}, \citenamefont {Kondor},\ and\
  \citenamefont {Cs{\'a}nyi}(2013)}]{bartok2013representing}%
  \BibitemOpen
  \bibfield  {author} {\bibinfo {author} {\bibfnamefont {A.~P.}\ \bibnamefont
  {Bart{\'o}k}}, \bibinfo {author} {\bibfnamefont {R.}~\bibnamefont {Kondor}},
  \ and\ \bibinfo {author} {\bibfnamefont {G.}~\bibnamefont {Cs{\'a}nyi}},\
  }\href@noop {} {\bibfield  {journal} {\bibinfo  {journal} {Phys. Rev. B}\
  }\textbf {\bibinfo {volume} {87}},\ \bibinfo {pages} {184115} (\bibinfo
  {year} {2013})}\BibitemShut {NoStop}%
\bibitem [{\citenamefont {El-Batanouny}\ and\ \citenamefont
  {Wooten}(2008)}]{el-batanouny_wooten_2008}%
  \BibitemOpen
  \bibfield  {author} {\bibinfo {author} {\bibfnamefont {M.}~\bibnamefont
  {El-Batanouny}}\ and\ \bibinfo {author} {\bibfnamefont {F.}~\bibnamefont
  {Wooten}},\ }\href {\doibase 10.1017/CBO9780511755736} {\emph {\bibinfo
  {title} {Symmetry and Condensed Matter Physics: A Computational Approach}}}\
  (\bibinfo  {publisher} {Cambridge University Press},\ \bibinfo {year}
  {2008})\BibitemShut {NoStop}%
\bibitem [{\citenamefont {{Tol{\'e}dano}}\ and\ \citenamefont
  {{Tol{\'e}dano}}(1987)}]{1987ltpt.book}%
  \BibitemOpen
  \bibfield  {author} {\bibinfo {author} {\bibfnamefont {J.-C.}\ \bibnamefont
  {{Tol{\'e}dano}}}\ and\ \bibinfo {author} {\bibfnamefont {P.}~\bibnamefont
  {{Tol{\'e}dano}}},\ }\href {\doibase 10.1142/0215} {\emph {\bibinfo {title}
  {{The Landau Theory of Phase Transitions: Application to Structural,
  Incommensurate, Magnetic and Liquid Crystal Systems}}}}\ (\bibinfo
  {publisher} {World Scientific Press},\ \bibinfo {year} {1987})\BibitemShut
  {NoStop}%
\bibitem [{\citenamefont {Hamermesh}(1989)}]{Hamermesh:1123140}%
  \BibitemOpen
  \bibfield  {author} {\bibinfo {author} {\bibfnamefont {M.}~\bibnamefont
  {Hamermesh}},\ }\href@noop {} {\emph {\bibinfo {title} {{Group Theory and its
  Application to Physical Problems}}}}\ (\bibinfo  {publisher} {Dover},\
  \bibinfo {address} {New York},\ \bibinfo {year} {1989})\BibitemShut {NoStop}%
\bibitem [{\citenamefont {Heine}(2007)}]{heine2007group}%
  \BibitemOpen
  \bibfield  {author} {\bibinfo {author} {\bibfnamefont {V.}~\bibnamefont
  {Heine}},\ }\href {https://books.google.co.jp/books?id=5WWqJnw0ySYC} {\emph
  {\bibinfo {title} {Group Theory in Quantum Mechanics: An Introduction to Its
  Present Usage}}},\ Dover Books on Physics\ (\bibinfo  {publisher} {Dover
  Publications},\ \bibinfo {year} {2007})\BibitemShut {NoStop}%
\bibitem [{\citenamefont {Chaichian}\ and\ \citenamefont
  {Hagedorn}(1997)}]{chaichian1997symmetries}%
  \BibitemOpen
  \bibfield  {author} {\bibinfo {author} {\bibfnamefont {M.}~\bibnamefont
  {Chaichian}}\ and\ \bibinfo {author} {\bibfnamefont {R.}~\bibnamefont
  {Hagedorn}},\ }\href {https://books.google.co.jp/books?id=pEhjQgAACAAJ}
  {\emph {\bibinfo {title} {Symmetries in Quantum Mechanics: From Angular
  Momentum to Supersymmetry (PBK)}}},\ Graduate Student Series in Physics\
  (\bibinfo  {publisher} {Taylor \& Francis},\ \bibinfo {year}
  {1997})\BibitemShut {NoStop}%
\bibitem [{oei()}]{oeis}%
  \BibitemOpen
  \href@noop {} {\enquote {\bibinfo {title} {{\rm The On-Line Encyclopedia of
  Integer Sequences}},}\ }\bibinfo {howpublished} {\url{http://oeis.org}},\
  \bibinfo {note} {{\rm OEIS} Foundation Inc. (2018)}\BibitemShut {NoStop}%
\bibitem [{\citenamefont {Kondor}(2008)}]{kondor2008group}%
  \BibitemOpen
  \bibfield  {author} {\bibinfo {author} {\bibfnamefont {I.~R.}\ \bibnamefont
  {Kondor}},\ }\href@noop {} {\emph {\bibinfo {title} {Group Theoretical
  Methods in Machine Learning}}}\ (\bibinfo  {publisher} {Columbia
  University},\ \bibinfo {year} {2008})\BibitemShut {NoStop}%
\bibitem [{\citenamefont {Steinhardt}, \citenamefont {Nelson},\ and\
  \citenamefont {Ronchetti}(1983)}]{PhysRevB.28.784}%
  \BibitemOpen
  \bibfield  {author} {\bibinfo {author} {\bibfnamefont {P.~J.}\ \bibnamefont
  {Steinhardt}}, \bibinfo {author} {\bibfnamefont {D.~R.}\ \bibnamefont
  {Nelson}}, \ and\ \bibinfo {author} {\bibfnamefont {M.}~\bibnamefont
  {Ronchetti}},\ }\href {\doibase 10.1103/PhysRevB.28.784} {\bibfield
  {journal} {\bibinfo  {journal} {Phys. Rev. B}\ }\textbf {\bibinfo {volume}
  {28}},\ \bibinfo {pages} {784} (\bibinfo {year} {1983})}\BibitemShut
  {NoStop}%
\bibitem [{\citenamefont {Carlsson}(1990)}]{carlsson1990beyond}%
  \BibitemOpen
  \bibfield  {author} {\bibinfo {author} {\bibfnamefont {A.~E.}\ \bibnamefont
  {Carlsson}},\ }in\ \href@noop {} {\emph {\bibinfo {booktitle} {Solid State
  Physics}}},\ Vol.~\bibinfo {volume} {43}\ (\bibinfo  {publisher} {Academic
  Press, Boston},\ \bibinfo {year} {1990})\ pp.\ \bibinfo {pages}
  {1--91}\BibitemShut {NoStop}%
\bibitem [{\citenamefont {Wood}\ and\ \citenamefont
  {Thompson}(2018{\natexlab{b}})}]{doi:10.1063/1.5017641}%
  \BibitemOpen
  \bibfield  {author} {\bibinfo {author} {\bibfnamefont {M.~A.}\ \bibnamefont
  {Wood}}\ and\ \bibinfo {author} {\bibfnamefont {A.~P.}\ \bibnamefont
  {Thompson}},\ }\href {\doibase 10.1063/1.5017641} {\bibfield  {journal}
  {\bibinfo  {journal} {J. Chem. Phys.}\ }\textbf {\bibinfo {volume} {148}},\
  \bibinfo {pages} {241721} (\bibinfo {year} {2018}{\natexlab{b}})}\BibitemShut
  {NoStop}%
\bibitem [{\citenamefont {Bergerhoff}\ and\ \citenamefont
  {Brown}(1987)}]{bergerhoff1987crystal}%
  \BibitemOpen
  \bibfield  {author} {\bibinfo {author} {\bibfnamefont {G.}~\bibnamefont
  {Bergerhoff}}\ and\ \bibinfo {author} {\bibfnamefont {I.~D.}\ \bibnamefont
  {Brown}},\ }in\ \href@noop {} {\emph {\bibinfo {booktitle} {Crystallographic
  Databases}}},\ \bibinfo {editor} {edited by\ \bibinfo {editor} {\bibfnamefont
  {F.~H.}\ \bibnamefont {Allen~et al.}}}\ (\bibinfo  {publisher} {International
  Union of Crystallography, Chester},\ \bibinfo {year} {1987})\BibitemShut
  {NoStop}%
\bibitem [{\citenamefont {Bl{\"o}chl}(1994)}]{PAW1}%
  \BibitemOpen
  \bibfield  {author} {\bibinfo {author} {\bibfnamefont {P.~E.}\ \bibnamefont
  {Bl{\"o}chl}},\ }\href@noop {} {\bibfield  {journal} {\bibinfo  {journal}
  {Phys. Rev. B}\ }\textbf {\bibinfo {volume} {50}},\ \bibinfo {pages} {17953}
  (\bibinfo {year} {1994})}\BibitemShut {NoStop}%
\bibitem [{\citenamefont {Perdew}, \citenamefont {Burke},\ and\ \citenamefont
  {Ernzerhof}(1996)}]{GGA:PBE96}%
  \BibitemOpen
  \bibfield  {author} {\bibinfo {author} {\bibfnamefont {J.~P.}\ \bibnamefont
  {Perdew}}, \bibinfo {author} {\bibfnamefont {K.}~\bibnamefont {Burke}}, \
  and\ \bibinfo {author} {\bibfnamefont {M.}~\bibnamefont {Ernzerhof}},\
  }\href@noop {} {\bibfield  {journal} {\bibinfo  {journal} {Phys. Rev. Lett.}\
  }\textbf {\bibinfo {volume} {77}},\ \bibinfo {pages} {3865} (\bibinfo {year}
  {1996})}\BibitemShut {NoStop}%
\bibitem [{\citenamefont {Kresse}\ and\ \citenamefont {Hafner}(1993)}]{VASP1}%
  \BibitemOpen
  \bibfield  {author} {\bibinfo {author} {\bibfnamefont {G.}~\bibnamefont
  {Kresse}}\ and\ \bibinfo {author} {\bibfnamefont {J.}~\bibnamefont
  {Hafner}},\ }\href@noop {} {\bibfield  {journal} {\bibinfo  {journal} {Phys.
  Rev. B}\ }\textbf {\bibinfo {volume} {47}},\ \bibinfo {pages} {558} (\bibinfo
  {year} {1993})}\BibitemShut {NoStop}%
\bibitem [{\citenamefont {Kresse}\ and\ \citenamefont
  {Furthm{\"u}ller}(1996)}]{VASP2}%
  \BibitemOpen
  \bibfield  {author} {\bibinfo {author} {\bibfnamefont {G.}~\bibnamefont
  {Kresse}}\ and\ \bibinfo {author} {\bibfnamefont {J.}~\bibnamefont
  {Furthm{\"u}ller}},\ }\href@noop {} {\bibfield  {journal} {\bibinfo
  {journal} {Phys. Rev. B}\ }\textbf {\bibinfo {volume} {54}},\ \bibinfo
  {pages} {11169} (\bibinfo {year} {1996})}\BibitemShut {NoStop}%
\bibitem [{\citenamefont {Kresse}\ and\ \citenamefont {Joubert}(1999)}]{PAW2}%
  \BibitemOpen
  \bibfield  {author} {\bibinfo {author} {\bibfnamefont {G.}~\bibnamefont
  {Kresse}}\ and\ \bibinfo {author} {\bibfnamefont {D.}~\bibnamefont
  {Joubert}},\ }\href@noop {} {\bibfield  {journal} {\bibinfo  {journal} {Phys.
  Rev. B}\ }\textbf {\bibinfo {volume} {59}},\ \bibinfo {pages} {1758}
  (\bibinfo {year} {1999})}\BibitemShut {NoStop}%
\bibitem [{\citenamefont {Tibshirani}(1996)}]{tibshirani1996regression}%
  \BibitemOpen
  \bibfield  {author} {\bibinfo {author} {\bibfnamefont {R.}~\bibnamefont
  {Tibshirani}},\ }\href@noop {} {\bibfield  {journal} {\bibinfo  {journal} {J.
  R. Stat. Soc. B}\ }\textbf {\bibinfo {volume} {58}},\ \bibinfo {pages} {267}
  (\bibinfo {year} {1996})}\BibitemShut {NoStop}%
\bibitem [{\citenamefont {Hastie}, \citenamefont {Tibshirani},\ and\
  \citenamefont {Friedman}(2009)}]{hastieelements}%
  \BibitemOpen
  \bibfield  {author} {\bibinfo {author} {\bibfnamefont {T.}~\bibnamefont
  {Hastie}}, \bibinfo {author} {\bibfnamefont {R.}~\bibnamefont {Tibshirani}},
  \ and\ \bibinfo {author} {\bibfnamefont {J.}~\bibnamefont {Friedman}},\
  }\href@noop {} {\emph {\bibinfo {title} {The Elements of Statistical
  Learning}}},\ \bibinfo {edition} {2nd}\ ed.\ (\bibinfo  {publisher}
  {Springer, New York},\ \bibinfo {year} {2009})\BibitemShut {NoStop}%
\bibitem [{\citenamefont {Strang}(2019)}]{strang2019linear}%
  \BibitemOpen
  \bibfield  {author} {\bibinfo {author} {\bibfnamefont {G.}~\bibnamefont
  {Strang}},\ }\href@noop {} {\emph {\bibinfo {title} {Linear Algebra and
  Learning from Data}}}\ (\bibinfo  {publisher} {Wellesley-Cambridge Press},\
  \bibinfo {year} {2019})\BibitemShut {NoStop}%
\bibitem [{\citenamefont {Branke}\ \emph {et~al.}(2008)\citenamefont {Branke},
  \citenamefont {Deb}, \citenamefont {Miettinen},\ and\ \citenamefont
  {Slowi{\'n}ski}}]{branke2008multiobjective}%
  \BibitemOpen
  \bibfield  {author} {\bibinfo {author} {\bibfnamefont {J.}~\bibnamefont
  {Branke}}, \bibinfo {author} {\bibfnamefont {K.}~\bibnamefont {Deb}},
  \bibinfo {author} {\bibfnamefont {K.}~\bibnamefont {Miettinen}}, \ and\
  \bibinfo {author} {\bibfnamefont {R.}~\bibnamefont {Slowi{\'n}ski}},\
  }\href@noop {} {\emph {\bibinfo {title} {Multiobjective Optimization:
  Interactive and Evolutionary Approaches}}},\ Vol.\ \bibinfo {volume} {5252}\
  (\bibinfo  {publisher} {Springer Science \& Business Media},\ \bibinfo {year}
  {2008})\BibitemShut {NoStop}%
\bibitem [{\citenamefont {Takahashi}, \citenamefont {Seko},\ and\ \citenamefont
  {Tanaka}(2018)}]{doi:10.1063/1.5027283}%
  \BibitemOpen
  \bibfield  {author} {\bibinfo {author} {\bibfnamefont {A.}~\bibnamefont
  {Takahashi}}, \bibinfo {author} {\bibfnamefont {A.}~\bibnamefont {Seko}}, \
  and\ \bibinfo {author} {\bibfnamefont {I.}~\bibnamefont {Tanaka}},\
  }\href@noop {} {\bibfield  {journal} {\bibinfo  {journal} {J. Chem. Phys.}\
  }\textbf {\bibinfo {volume} {148}},\ \bibinfo {pages} {234106} (\bibinfo
  {year} {2018})}\BibitemShut {NoStop}%
\bibitem [{\citenamefont {Zope}\ and\ \citenamefont
  {Mishin}(2003)}]{PhysRevB.68.024102}%
  \BibitemOpen
  \bibfield  {author} {\bibinfo {author} {\bibfnamefont {R.~R.}\ \bibnamefont
  {Zope}}\ and\ \bibinfo {author} {\bibfnamefont {Y.}~\bibnamefont {Mishin}},\
  }\href {\doibase 10.1103/PhysRevB.68.024102} {\bibfield  {journal} {\bibinfo
  {journal} {Phys. Rev. B}\ }\textbf {\bibinfo {volume} {68}},\ \bibinfo
  {pages} {024102} (\bibinfo {year} {2003})}\BibitemShut {NoStop}%
\bibitem [{\citenamefont {Fujii}\ and\ \citenamefont
  {Seko}(2022)}]{FUJII2022111137}%
  \BibitemOpen
  \bibfield  {author} {\bibinfo {author} {\bibfnamefont {S.}~\bibnamefont
  {Fujii}}\ and\ \bibinfo {author} {\bibfnamefont {A.}~\bibnamefont {Seko}},\
  }\href {\doibase https://doi.org/10.1016/j.commatsci.2021.111137} {\bibfield
  {journal} {\bibinfo  {journal} {Comput. Mater. Sci.}\ }\textbf {\bibinfo
  {volume} {204}},\ \bibinfo {pages} {111137} (\bibinfo {year}
  {2022})}\BibitemShut {NoStop}%
\bibitem [{\citenamefont {Togo}\ and\ \citenamefont
  {Tanaka}(2015)}]{PhonopyArticle}%
  \BibitemOpen
  \bibfield  {author} {\bibinfo {author} {\bibfnamefont {A.}~\bibnamefont
  {Togo}}\ and\ \bibinfo {author} {\bibfnamefont {I.}~\bibnamefont {Tanaka}},\
  }\href@noop {} {\bibfield  {journal} {\bibinfo  {journal} {Scr. Mater.}\
  }\textbf {\bibinfo {volume} {108}},\ \bibinfo {pages} {1} (\bibinfo {year}
  {2015})}\BibitemShut {NoStop}%
\bibitem [{\citenamefont {Togo}, \citenamefont {Chaput},\ and\ \citenamefont
  {Tanaka}(2015)}]{PhysRevB.91.094306}%
  \BibitemOpen
  \bibfield  {author} {\bibinfo {author} {\bibfnamefont {A.}~\bibnamefont
  {Togo}}, \bibinfo {author} {\bibfnamefont {L.}~\bibnamefont {Chaput}}, \ and\
  \bibinfo {author} {\bibfnamefont {I.}~\bibnamefont {Tanaka}},\ }\href
  {\doibase 10.1103/PhysRevB.91.094306} {\bibfield  {journal} {\bibinfo
  {journal} {Phys. Rev. B}\ }\textbf {\bibinfo {volume} {91}},\ \bibinfo
  {pages} {094306} (\bibinfo {year} {2015})}\BibitemShut {NoStop}%
\bibitem [{\citenamefont {Nishiyama}, \citenamefont {Seko},\ and\ \citenamefont
  {Tanaka}(2020)}]{PhysRevMaterials.4.123607}%
  \BibitemOpen
  \bibfield  {author} {\bibinfo {author} {\bibfnamefont {T.}~\bibnamefont
  {Nishiyama}}, \bibinfo {author} {\bibfnamefont {A.}~\bibnamefont {Seko}}, \
  and\ \bibinfo {author} {\bibfnamefont {I.}~\bibnamefont {Tanaka}},\ }\href
  {\doibase 10.1103/PhysRevMaterials.4.123607} {\bibfield  {journal} {\bibinfo
  {journal} {Phys. Rev. Mater.}\ }\textbf {\bibinfo {volume} {4}},\ \bibinfo
  {pages} {123607} (\bibinfo {year} {2020})}\BibitemShut {NoStop}%
\bibitem [{\citenamefont {Stillinger}\ and\ \citenamefont
  {Weber}(1985)}]{PhysRevB.31.5262}%
  \BibitemOpen
  \bibfield  {author} {\bibinfo {author} {\bibfnamefont {F.~H.}\ \bibnamefont
  {Stillinger}}\ and\ \bibinfo {author} {\bibfnamefont {T.~A.}\ \bibnamefont
  {Weber}},\ }\href {\doibase 10.1103/PhysRevB.31.5262} {\bibfield  {journal}
  {\bibinfo  {journal} {Phys. Rev. B}\ }\textbf {\bibinfo {volume} {31}},\
  \bibinfo {pages} {5262} (\bibinfo {year} {1985})}\BibitemShut {NoStop}%
\bibitem [{\citenamefont {Tersoff}(1988)}]{PhysRevB.37.6991}%
  \BibitemOpen
  \bibfield  {author} {\bibinfo {author} {\bibfnamefont {J.}~\bibnamefont
  {Tersoff}},\ }\href {\doibase 10.1103/PhysRevB.37.6991} {\bibfield  {journal}
  {\bibinfo  {journal} {Phys. Rev. B}\ }\textbf {\bibinfo {volume} {37}},\
  \bibinfo {pages} {6991} (\bibinfo {year} {1988})}\BibitemShut {NoStop}%
\bibitem [{\citenamefont {Ackland}\ \emph {et~al.}(1987)\citenamefont
  {Ackland}, \citenamefont {Tichy}, \citenamefont {Vitek},\ and\ \citenamefont
  {Finnis}}]{doi:10.1080/01418618708204485}%
  \BibitemOpen
  \bibfield  {author} {\bibinfo {author} {\bibfnamefont {G.~J.}\ \bibnamefont
  {Ackland}}, \bibinfo {author} {\bibfnamefont {G.}~\bibnamefont {Tichy}},
  \bibinfo {author} {\bibfnamefont {V.}~\bibnamefont {Vitek}}, \ and\ \bibinfo
  {author} {\bibfnamefont {M.~W.}\ \bibnamefont {Finnis}},\ }\href {\doibase
  10.1080/01418618708204485} {\bibfield  {journal} {\bibinfo  {journal}
  {Philos. Mag. A}\ }\textbf {\bibinfo {volume} {56}},\ \bibinfo {pages} {735}
  (\bibinfo {year} {1987})}\BibitemShut {NoStop}%
\bibitem [{\citenamefont {Williams}, \citenamefont {Mishin},\ and\
  \citenamefont {Hamilton}(2006)}]{Williams_2006}%
  \BibitemOpen
  \bibfield  {author} {\bibinfo {author} {\bibfnamefont {P.~L.}\ \bibnamefont
  {Williams}}, \bibinfo {author} {\bibfnamefont {Y.}~\bibnamefont {Mishin}}, \
  and\ \bibinfo {author} {\bibfnamefont {J.~C.}\ \bibnamefont {Hamilton}},\
  }\href {\doibase 10.1088/0965-0393/14/5/002} {\bibfield  {journal} {\bibinfo
  {journal} {Model. Simul. Mater. Sci. Eng.}\ }\textbf {\bibinfo {volume}
  {14}},\ \bibinfo {pages} {817} (\bibinfo {year} {2006})}\BibitemShut
  {NoStop}%
\bibitem [{\citenamefont {Zhou}, \citenamefont {Johnson},\ and\ \citenamefont
  {Wadley}(2004)}]{PhysRevB.69.144113}%
  \BibitemOpen
  \bibfield  {author} {\bibinfo {author} {\bibfnamefont {X.~W.}\ \bibnamefont
  {Zhou}}, \bibinfo {author} {\bibfnamefont {R.~A.}\ \bibnamefont {Johnson}}, \
  and\ \bibinfo {author} {\bibfnamefont {H.~N.~G.}\ \bibnamefont {Wadley}},\
  }\href {\doibase 10.1103/PhysRevB.69.144113} {\bibfield  {journal} {\bibinfo
  {journal} {Phys. Rev. B}\ }\textbf {\bibinfo {volume} {69}},\ \bibinfo
  {pages} {144113} (\bibinfo {year} {2004})}\BibitemShut {NoStop}%
\end{thebibliography}%

\end{document}